\documentclass{aastex}
\usepackage{emulateapj5}
\pdfoutput=1

\usepackage{natbib}                    

\shortauthors{Chiboucas et al.}
\shorttitle{UCDs in Coma}

\begin{document}

\title{Ultra-Compact Dwarfs in the Coma Cluster }

\author{Kristin Chiboucas\altaffilmark{1},  R. Brent Tully\altaffilmark{2}, R. O. Marzke\altaffilmark{3}, S. Phillipps\altaffilmark{4}, J. Price\altaffilmark{4}, E. Peng\altaffilmark{5}, Neil Trentham \altaffilmark{6}, David Carter\altaffilmark{7}, and Derek Hammer\altaffilmark{8}}
\email{kchibouc@gemini.edu, tully@ifa.hawaii.edu, marzke@sfsu.edu,S.Phillipps@bristol.ac.uk,james.je.price@gmail.com,peng@bac.pku.edu.cn,trentham@ast.cam.ac.uk, dxc@astro.livjm.ac.uk, hammerd@pha.jhu.edu}

\altaffiltext{1}{Gemini Observatory, 670 N. A'ohoku Pl, Hilo, HI 96720, USA}
\altaffiltext{2}{Institute for Astronomy, University of Hawaii, 2680 Woodlawn Dr., Honolulu, HI 96821, USA}
\altaffiltext{3}{Department of Physics and Astronomy, San Francisco State University, San Francisco, CA 94132, USA}
\altaffiltext{4}{Astrophysics Group, H.H. Wills Physics Laboratory, University of Bristol, Tyndall Avenue, Bristol BS8 1TL, UK}
\altaffiltext{5}{Kavli Institute for Astronomy and Astrophysics, Peking University, Beijing 100871, China}
\altaffiltext{6}{Institute of Astronomy, Madingley Road, Cambridge CB3 0HA, UK}
\altaffiltext{7}{Astrophysics Research Institute, Liverpool John Moores University, Twelve Quays House, Egerton Wharf, Birkenhead CH411LD, UK}
\altaffiltext{8}{Department of Physics and Astronomy, Johns Hopkins University, 3400 North Charles Street, Baltimore, MD 21218}

\begin{abstract}

We have undertaken a spectroscopic search for ultra compact dwarf galaxies (UCDs)
in the dense core of the dynamically evolved, massive Coma cluster as part of the
HST/ACS Coma Cluster Treasury Survey.  UCD candidates
were initially chosen based on color, magnitude, degree of resolution within the ACS
images, and the known properties of Fornax and Virgo UCDs.
Follow-up spectroscopy with Keck/LRIS confirmed 27 candidates as members of the 
Coma Cluster, a success rate $> 60\%$ for targeted 
objects brighter than M$_R = -12$.  Another 14 candidates may also prove to be Coma 
members, but low signal-to-noise spectra prevent definitive conclusions.  
An investigation of the properties and distribution of the Coma UCDs finds these 
objects to be very similar to UCDs discovered in other environments.   
The Coma UCDs tend to be clustered around giant galaxies in the cluster core and have 
colors/metallicity that correlate with the host galaxy.
With properties and a distribution similar to that of the 
Coma cluster globular cluster population, we find strong support for a star cluster 
origin for the majority of the Coma UCDs.  However, a few UCDs appear to have 
stellar population or structural properties which differentiate them from
the old star cluster populations found in the Coma cluster, perhaps indicating that 
UCDs may form through multiple formation channels.
                                                                
\end{abstract}


\section{Introduction}\label{intro}

Spectroscopic redshifts provide the most reliable method for establishing
membership in galaxy clusters, but are prohibitively time consuming
considering the abundance of faint objects in cluster fields. 
Given this limitation, studies of faint cluster galaxies often use indirect
means such as selecting probable members by color, surface brightness, and
morphology.  Such indirect methods, however, require assumptions about the 
properties of cluster members.
Consequently, these methods are inherently biased. Entire
populations of galaxies may be left out of cluster member samples.

It had previously been assumed that compact high surface brightness
objects were either
background ellipticals or foreground stars.  This was due in part to
a well defined surface brightness-magnitude relation 
in which dwarf elliptical galaxies trend toward lower surface brightnesses
at fainter magnitudes \citep{cb87,fs88,ibm88,bc91,mieske04b}. 
Only a decade ago,
spectroscopic surveys of the Fornax cluster, including an all-object 2dF 
spectroscopic survey, revealed a new class of object termed ultra-compact
dwarf galaxies (UCDs) \citep{hil99, djgp00, phill01}.  These faint objects
are offset from the magnitude-surface brightness relation of dE galaxies
having high surface brightnesses and very small sizes.

This discovery spurred dedicated searches for UCDs in other nearby groups
and clusters.  A large population of about 60 UCDs has since been established to 
exist in Fornax \citep{goud01,drink05,firth08,gregg09}.  A similarly large population
has recently been identified in the Hydra cluster \citep{misgeld11}, while
somewhat smaller populations have been found in the core 
of the Virgo ($\sim 25$ UCDs) \citep{has05,jones06,firth08}  and Centaurus 
clusters \citep{mieske07,mieske09}.  A single UCD was confirmed in
the Dorado group \citep{Ev07b}.  
Candidate UCDs have also been identified in
the Coma, Hydra, A1689, and AS0740 clusters and NGC1023 group
\citep{adamiucd,juan,wehner07,mieske04,blake08,mieske06}.  Only a handful of UCDs have been
detected in low-density environments, either as companions to isolated field galaxies
or in poor groups \citep{hau09, darocha10, norris11}.
None have been found in the poor Local or nearby M81 Group \citep{kcm81}
environments.

UCDs are intrinsically faint and compact objects that are unresolved in 
ground-based surveys and have magnitudes between $-13.5 < M_V < -10.5$.
Effective radii for these objects span the range $7 <$ r$_e < 100$ pc
\citep{mieske07,jones06,mieskeb08},  which is
larger than typical globular clusters with sizes
$\sim 2-5$ pc \citep{larsen01a,jordan05} but smaller than dwarf ellipticals which have sizes of a few 
100 parsecs \citep{drink03}.  At the distance
of the Coma cluster, the UCD half-light radii range corresponds to $0.01 - 0.2$ arcsec.
Thus, with the Advanced Camera for Surveys (ACS) $\sim0.1$ arcsec resolution, the larger 
ones are just resolved.  
The brighter UCDs also have exceptionally red optical colors, lying redward of the red 
sequence by up to $\sim0.2$ magnitudes
in Virgo and Fornax, redder than typical globular clusters by 0.1 magnitudes,
and redder than the nuclei of dE,N galaxies \citep{mieske06b}.  The red color may be
driven by high metallicities \citep{mieske06b,chil08}.  With these distinct properties,
it is clear that UCDs constitute a separate class of object from the abundant dE and dSph
galaxies found in clusters and groups.

Measurements of mass-to-light (M/L) ratios have been made for a few of these objects and
found to range from $2-9$ \citep{mieskeb08,has05,Ev07a,hil07}.  The 
larger values 
could be evidence for the presence of dark matter but alternative 
explanations include: (1) inflated mass estimates produced by tidal heating which increases
the velocity dispersion \citep{fk06} or (2) bottom  
\citep{mieske08c} or top heavy IMFs \citep{dab09,murray09}.

A number of potential formation mechanisms have been proposed to
explain the nature and origin of these enigmatic objects 
\citep{djgp00,fk02,bekki03,has05,chil08,Ev08}.  UCDs may simply be globular star clusters
comprising the extreme bright tail of the globular cluster luminosity function or
super star cluster end products from the
mergers of massive star clusters themselves formed during galaxy mergers.
Alternatively, they may be the visible
nuclei of dwarf ellipticals with exceptionally low surface brightness
envelopes, primordial objects formed of the small scale peaks 
in the initial power law spectrum, or remnant nuclei of tidally stripped 
nucleated dwarf elliptical or late type galaxies which have been 
"threshed" during tidal encounters with 
giant galaxies and with the cluster potential.  While this latter 
explanation is sometimes favored, other possibilities
have not been ruled out.  In particular, evidence exists
showing that UCDs may be more metal rich and, in some cases, older than the nuclei
of dE,N from which they purportedly originated \citep{chil08,mieske06b}.
Because dE,N can continue to form stars in their cores through gas accretion
while threshed nuclei, having lost their outer halo, suffer from strangulation
and a cessation of star formation, it is difficult to reconcile the higher
metallicities and, at the same time, older ages with a threshing mechanism.
These studies instead support a super star cluster scenario originating from the merger 
of young massive clusters at early times.  In addition, UCD sizes tend to be larger
than both globular clusters and the nuclei of dwarf galaxies \citep{dep05}.  
However, finding age and metallicity measurements for UCDs and dwarf nuclei in 
Virgo that are compatible,  
\citet{paudel10} argue that the stripping scenario is a viable option.  
Yet other researchers suggest that with a continuum of
luminosities and with colors similar to metal rich globular clusters,
UCDs may simply constitute a bright extension of the globular cluster sequence \citep{wehner07}.

If UCDs originate as remnant dE nuclei via a threshing mechanism they would be
expected to populate cores of massive and dynamically evolved clusters.
Compared to lower density environments, they would be expected in greater numbers 
and with a broader distribution within the massive Coma cluster.
Alternatively, if UCDs are giant globular clusters or 
super star clusters they should be associated with 
individual galaxies and have properties similar to those of the host galaxy's 
globular cluster and stellar populations.
Indeed, evidence has shown UCD populations in Fornax and Virgo are strongly
clustered, having smaller 
velocity dispersions than cluster dwarfs or even the central giant galaxy
globular cluster systems \citep{gregg09,mieske04b,jones06,firth08}. 
\citet{firth08} have furthermore found
that the majority of UCDs lie within the fields of the dominant giant galaxies
in Virgo and Fornax while very few have been discovered to lie in intracluster regions.
Therefore, we might expect to find a population of UCDs
within the Coma cluster, associated with either the cluster potential or individual
giant galaxies depending on formation mechanism.

This work is part of the larger HST/ACS Coma Cluster Treasury Survey
\citep{carter}, a two-passband imaging survey designed to cover 740 
arcmin$^{2}$ in the Coma Cluster to a depth of $I_C \sim 26.6$ mag for point
sources.  These
data are being used to perform comprehensive structural, photometric,
and morphological studies of the Coma Cluster members.  Only
28\% of the originally proposed areal coverage was completed due to the
failure of ACS, but it includes
much of the core region.  The goal of the Keck/Low-Resolution Imaging Spectrometer 
(LRIS) study
discussed in this work was to measure redshifts and establish membership
for galaxies at the faint-end of the cluster luminosity function.
We specifically targeted two samples of faint galaxies.  The first,
which is discussed in \citet{lris}, targeted galaxies in the
magnitude range $19 < R < 22  (-16 < M_R < -13)$ having low surface brightness and
membership previously estimated through indirect means.
The second sample, which we discuss here, targeted candidate
UCDs with $R < 24$ ($M_R < -11$). Candidates were chosen in the core
region of the Coma Cluster where HST/ACS data had already been obtained.

In Section \ref{obs} we describe the observations and data reduction procedure, in 
Section \ref{results} we detail the properties and distribution of confirmed cluster
member UCDs, and in Section \ref{disc} we discuss these results and potential 
origins.  A summary is presented in Section \ref{conc}. 
Throughout this paper, we assume a distance to the Coma Cluster of 100 Mpc 
and a distance modulus $\mu = 35$ \citep[see][Table 1]{carter}.

\section{Observations}\label{obs}

\subsection{Target Selection}\label{targs}

When this project was begun, the ACS data had just been taken and 
images were not yet calibrated.  
Therefore, to choose a sample of potential UCDs, we first identified point
sources in the catalog of \citet{adami} based on their large ground-based
Coma Cluster CFHT/CFH12K survey.  We then imposed the following criteria:
\begin{enumerate}
\item $R < 24$ (Vega mag).
\item $0.45 < (B-V) < 1.1$ ($3^{\prime\prime}$ apertures).
\item Higher priority to sources with $0.15 < R-I < 0.6$.
\item Located in the core region with processed ACS images available before 2007 Feb 11.
\end{enumerate}
The $B-V$ range includes the Fornax/Virgo UCDs at the blue end and
cEs at the red end.  In part because of large errors in the Adami
catalog at these faint magnitudes (the typical error in $B-V$ at $R = 23.5$ is
0.4 mag), we refrained from making our color cut too tight.

After the initial magnitude and color selection which generated a list of 
165 good candidates, a weight was given
to each object based on the deviation of the ACS image object profile from the ACS PSF.  
Higher weights were 
given to those objects with any sign of having a FWHM broader than stellar.
These weights were fed into the LRIS Autoslit3 mask-making software to be used in cases
of slit conflicts.  For astrometry accurate to $\sim 0.1$ arcsec,
we transformed the world coordinate system of the ACS images to the SDSS 
system.   Target coordinates were then measured from the ACS images.
In an initial 2008 observing run, a total of 47 UCD candidates were 
chosen by the software to populate 4 different masks.  A second sample of 
93 lower surface brightness
galaxies were observed concurrently with these 4 masks with the goal of 
establishing cluster membership 
for faint dwarf galaxies.  The results of this separate study are presented
in \citet{lris} and Trentham et al. (in prep).

The 2008 multi-object spectroscopy observations confirmed 19 Coma cluster member UCDs.
These original candidates were 
chosen based on the properties of known UCDs in the Virgo and Fornax clusters.
Because the full range of properties for these poorly understood objects is not well known,
we expanded our search during a 2009 campaign to more fully explore the
boundaries of the UCD parameter space.  Targets were chosen based on a broader
set of criteria, allowing targets with a wider range in color ($B-V < 1.2$) and
with a greater range of resolution, including more unresolved objects.
Where ACS imaging within the core region did not exist, we chose targets based strictly on
the ground-based photometry from \citet{adami}.  Greater weight
was given to brighter targets, in particular to explore a magnitude gap ($21.5 < R < 20.2$)
discovered to exist between compact elliptical galaxies and UCDs.  72 candidates were
chosen for 2 masks, from which a further 8 clusters members were identified.

\subsection{Observations and Data Reduction}\label{obsspec}
 
Observations and data reduction are described in \citet{lris}, but 
we summarize here.
Because we are observing faint targets, down to $R < 24$, along with low
surface brightness galaxies, the spectroscopic design was intended to 
maximize light throughput at
the expense of resolution.  As we intend only to measure redshifts in order to
distinguish between stars, Coma cluster members at $z=0.023$, and background objects,
and the Coma Cluster has a velocity
dispersion of $\sim1000$ km/s, a resolution of 200 km/s is 
adequate.   The spectral range was chosen to include the 4000\AA \ Balmer break 
and Ca H and K lines at the blue end ($\sim 4060$\AA \ at the redshift
of Coma) as these are some of the strongest spectral features in quiescent,  
low redshift, faint galaxies.

We use the Keck LRIS low resolution imaging spectrometer in multi-object spectroscopy mode  
which has very high sensitivity in the blue.  A dichroic was used to split light
at 5600\AA \ between red and blue chips.  On the blue-side, we used
the 400 line mm$^{-1}$ grism blazed at 3400\AA \ providing a dispersion of
1.07\AA \ pix$^{-1}$ and wavelength coverage of 4384\AA. With 1.2 arcsec
slitlets, we achieved a resolution of 7.8\AA \ FWHM. 
On the red-side we chose the 400 line mm$^{-1}$ grating blazed at 8500\AA \
with wavelength coverage 3950\AA \ and 1.92\AA \ pix$^{-1}$ dispersion.
The red-side data are used primarily to identify high redshift objects with emission
line spectra.

\begin{figure}[t]
\begin{centering}
\includegraphics[angle=0,totalheight=5.0in]{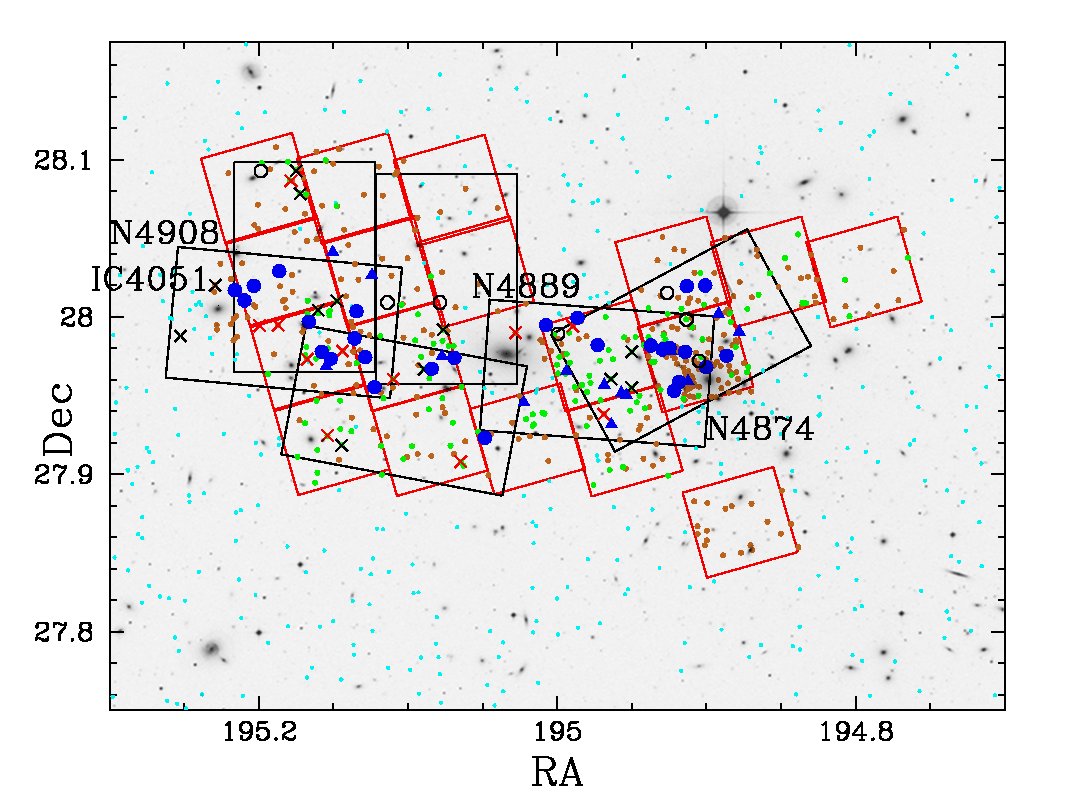}
\caption[]{Spatial distribution of the UCDs. Red boxes are the locations of observed 
ACS fields,
black rectangles are the 6 LRIS masks. Green points represent our original candidate sample, 
brown points are the expanded candidate sample, and cyan points are candidates chosen
strictly on the basis of color. 
Large solid circles denote the location of confirmed UCDs and triangles as less 
certain UCDs. Open circles mark the location of compact dEs \citep{compact}. X's are 
UCD candidates determined from redshifts to be stars (black) and 
background galaxies (red).
\label{overlay}}
\end{centering}
\end{figure}

Four masks were observed over 2 nights, 2-3 April 2008. Another 2 masks
were observed 28-30 March 2009 in poor conditions.  Each mask
covers a field of view of $5^{\prime} \times 8^{\prime}$ with an
average 35 slitlets.
Total integrations for each mask during the first run were $8 \times 1500$s with the exception
of the blue side of one mask for which we obtained $6 \times 1500$s.
During the second run we were plagued with thick clouds and very poor seeing.  Total
integrations were $\sim8 \times 1500$s but effective exposure times are much less.
Signal-to-noise $\AA^{-1}$ (measured around 5000\AA) for the UCD candidate spectra 
range from $\sim 20$ to less than 1. Secure redshifts were measured from spectra with 
S/N $> 4.0$, while less secure measurements came from $2.5 <$ S/N $< 5$
spectra.  Table \ref{obinfo} provides a summary of the observations.
In Figure \ref{overlay}, we overlay the locations of the completed ACS fields
and the 6 LRIS masks on an image of the central region of the Coma cluster.
The original and expanded set of candidate UCDs are shown as green and brown/cyan 
points, respectively.

Data were reduced using the standard procedures in IRAF.  Images
were overscan corrected and corrected for different gains. 
Halogen flats for each mask were combined and a normalized flat image was
generated with APFLATTEN.  Following division by this
flat, the 8 individual exposures for each mask were
median combined using sigma clipping.
Slit spectra were rectified by tracing the slit gaps for each mask and
fitting these with 4th order Legendre polynomials.  GEOMAP was run
to compute the two-dimensional surface for the full set of slit gaps in a mask and,
using that transformation, GEOTRAN was then executed to
generate images with straightened slits.
Arc spectra were rectified for each slitlet in the same manner.

The usual IRAF tasks IDENTIFY, FITCOORD, and TRANSFORM were used to
wavelength calibrate the object spectra from arc lamp spectra.  Since arc spectra were taken
only once per night, sky lines in each object 
spectra were used to correct for offsets from the lamp wavelength 
calibrations.  Unfortunately, the prominent 5577\AA \ sky line fell on the edge
of our spectra, or depending on the location in the mask, off the
observed blue-side spectra altogether.  Due to uncertainties in the
applied shift, we therefore expect systematic errors of up to 100 km/s
in the radial velocity measurements from our 2008 run targets.  Observations
taken in 2009 were bracketed with arc lamp exposures and are not expected
to suffer from these systematic offsets.

To extract one-dimensional spectra, we used APALL to identify the
spectrum center, width, and sky regions.  RVSAO/XCSAO was used,
along with absorption and emission line template spectra, to measure redshifts.
A flux standard was observed during a later run with the same configuration,
used for relative flux calibration.                                                                                     
\section{Results}\label{results}

\begin{figure}[t]
\begin{centering}
\plotone{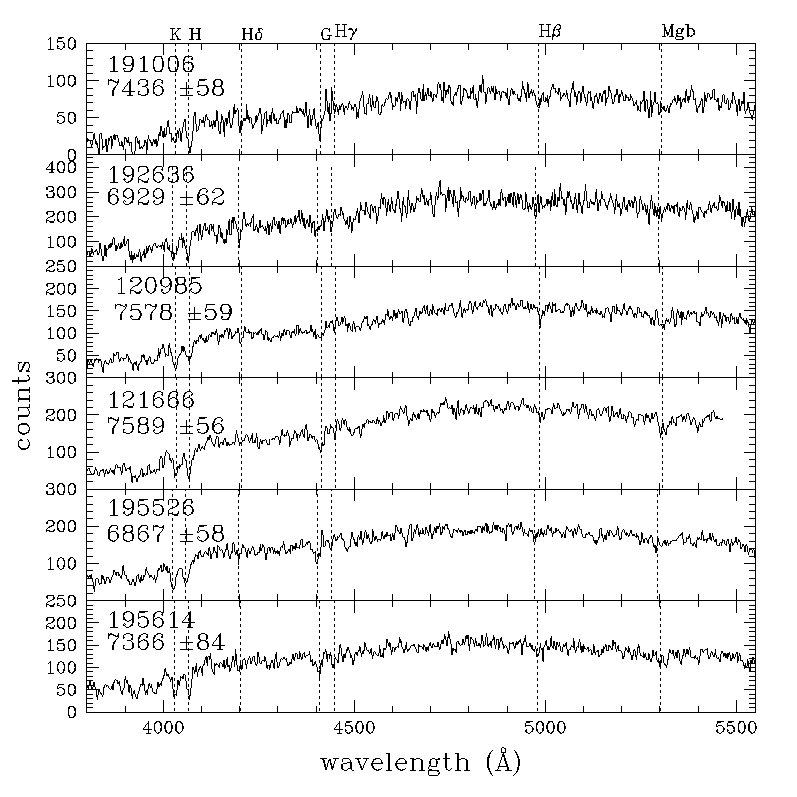}
\caption[]{Spectra for 6 of our UCDs, smoothed 3 times.  For each, we list the object ID as in 
Tables \ref{ucdtab}-\ref{ucdtab2}, along with the radial velocity in km s$^{-1}$ derived from absorption lines.
\label{spect1s}}
\end{centering}
\end{figure}

\begin{figure}[t]
\begin{centering}
\plotone{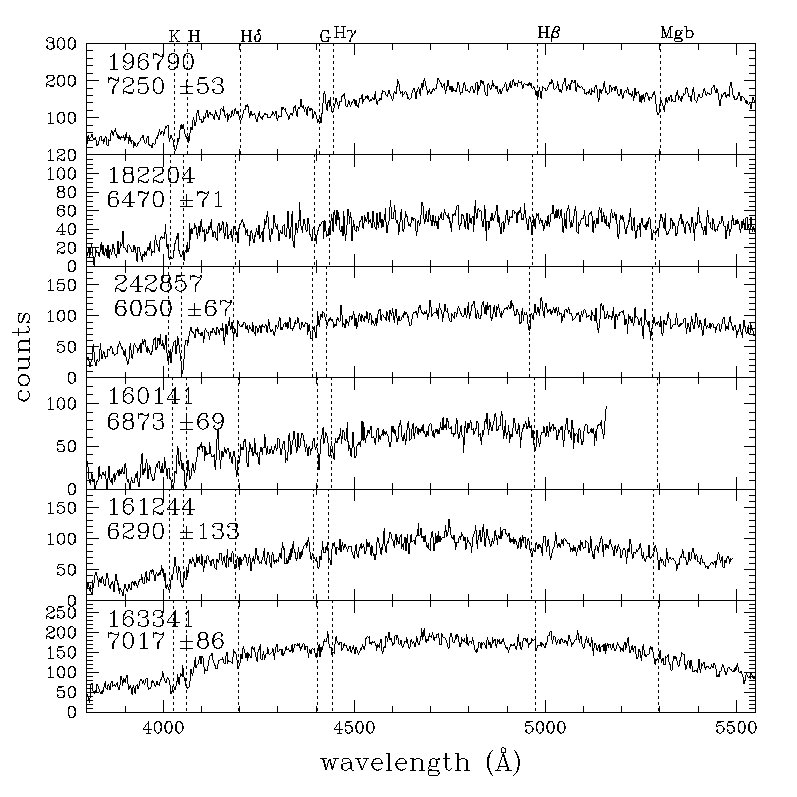}
\caption[]{Spectra for another 6 UCDs, smoothed 3 times. Labels as in Figure \ref{spect1s}.
\label{spect2s}}
\end{centering}
\end{figure}

\begin{figure}[t]
\begin{centering}
\plotone{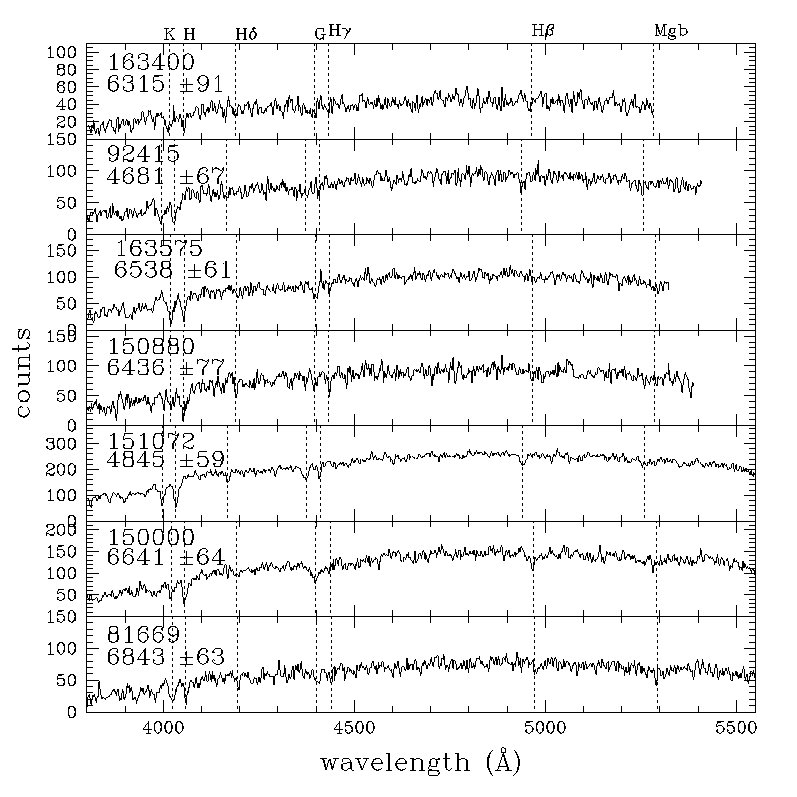}
\caption[]{Spectra for 7 confirmed UCDs, smoothed 3 times. Labels as in Figure \ref{spect1s}.
\label{spect3s}}
\end{centering}
\end{figure}

\begin{figure}[t]
\begin{centering}
\plotone{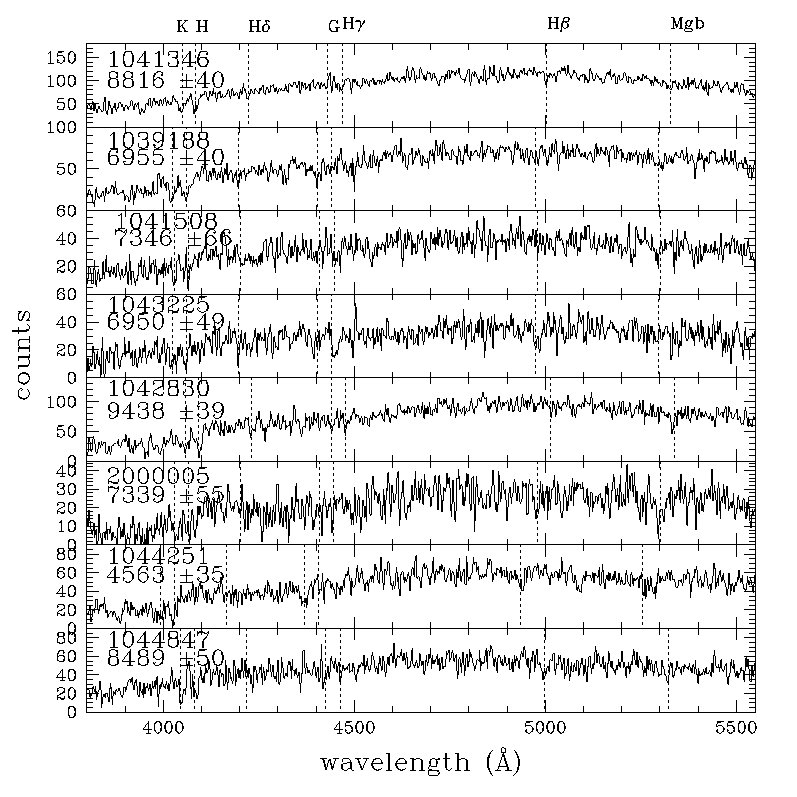}
\caption[]{Spectra for the second set of 8 confirmed UCDs, smoothed 3 times. Labels as in Figure \ref{spect1s}.
\label{spect4s}}
\end{centering}
\end{figure}

\begin{figure}[t]
\begin{centering}
\plotone{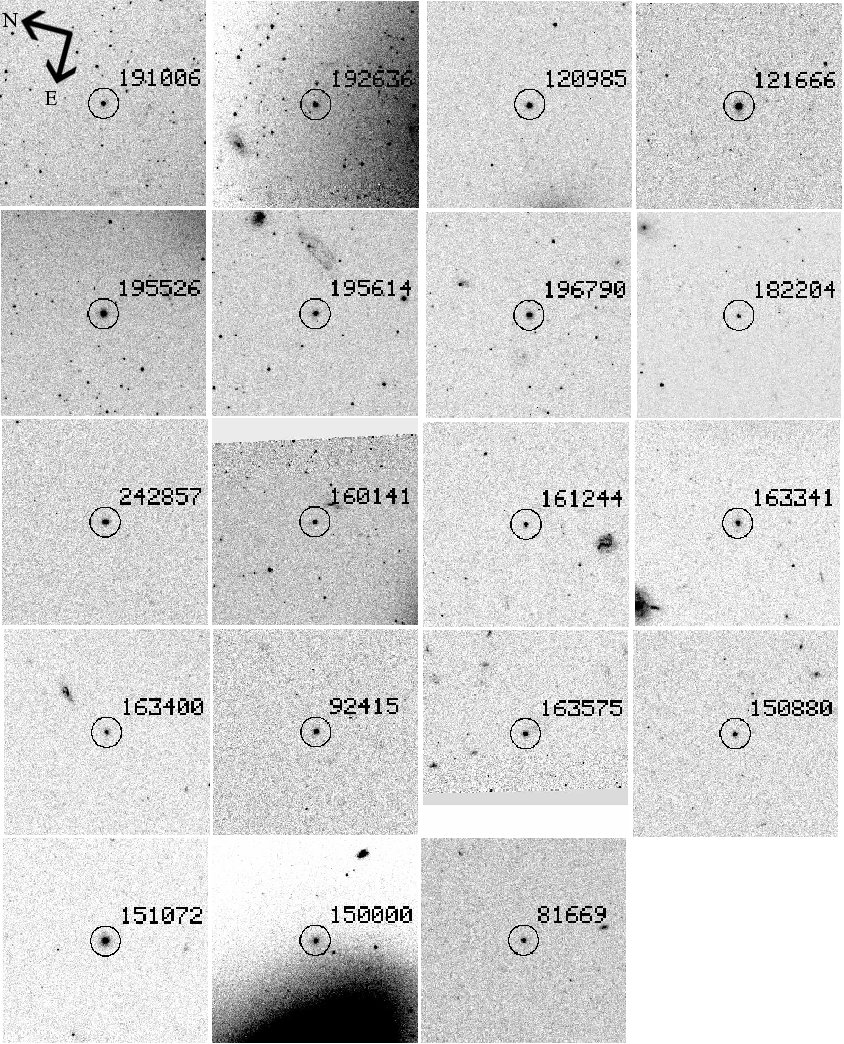}
\caption[]{ACS $F814W-$band thumbnail images, 17.5 arcsec across for all 19 confirmed UCDs.  
Circles that identify the UCDs are 2.6 arcsec in diameter.
\label{thumb}}
\end{centering}
\end{figure}

\begin{figure}[t]
\begin{centering}
\plotone{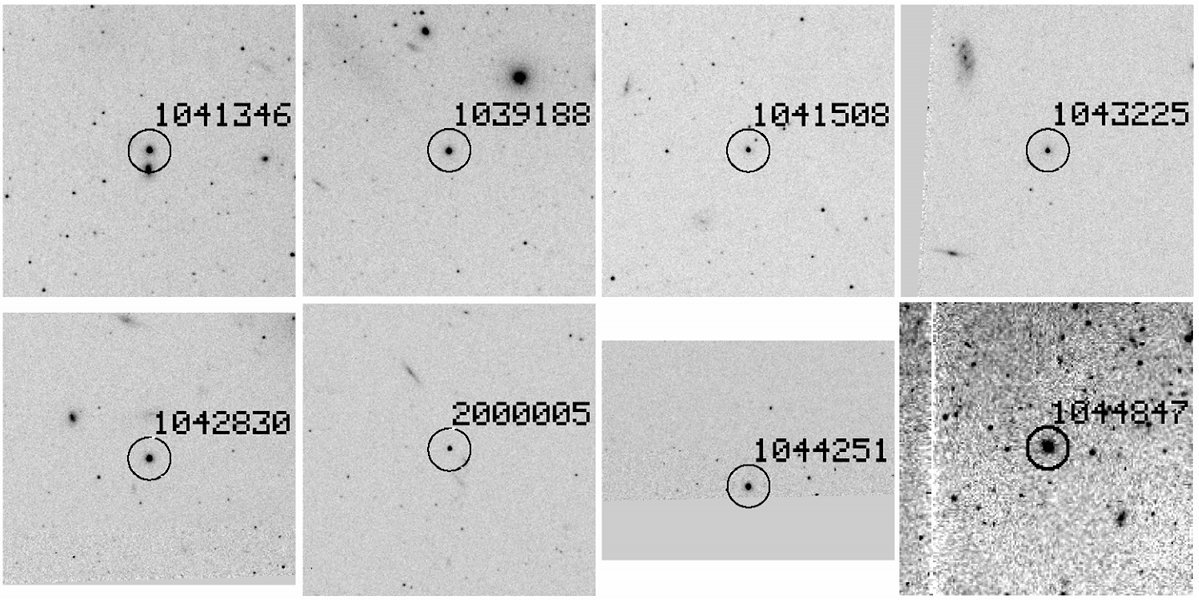}
\caption[]{ACS $F814W-$band thumbnail images, 17.5 arcsec across for 8 UCDs 
in the second sample. 
The WFPC2 image containing 1044847 comes from archival HST data (program GO6283, PI J. Westphal).
\label{thumb2}}
\end{centering}
\end{figure}

Tables \ref{ucdtab}-\ref{ucdtab2} list spectroscopic redshift measurements for the UCD 
candidate samples.   
We assume that objects with radial velocities between $4000 < v_r < 10000$ km s$^{-1}$,
within $3\sigma$ of the cluster mean, are cluster members.
In total, we find 27 compact sources with redshift measurements consistent
with Coma cluster measurement.  Another 14 also have measured redshifts
that would place them in the Coma cluster, but these come from very
low S/N ($ < 5$) spectra and we consider these measurements to 
be highly uncertain.  Out of the first set of 47 targeted candidates, 19
proved to be members, 6 were highly uncertain members, 4 turned out to be background
galaxies, and 6 proved to be foreground stars. The remaining
12 targets had spectra with too low S/N to even attempt redshift measurements.
Spectra for these 19 confirmed UCDs are shown in Figures \ref{spect1s} -
\ref{spect3s}.
Thumbnail images of these confirmed UCDs are provided in Fig \ref{thumb}.  
The second round of observations based on our expanded candidate sample
turned up 7 background galaxies and 6 stars along with 8 members.  Spectra for these, taken in 
poor conditions, are shown in Figure \ref{spect4s} and thumbnails are presented in 
Figure \ref{thumb2}. 
In Figure \ref{overlay}, we display the spatial distribution of all
confirmed and questionable UCDs along with the full set of candidates.
Spectroscopically determined stars and background galaxies are also highlighted.

\begin{figure}[t]
\begin{centering}
\plottwo{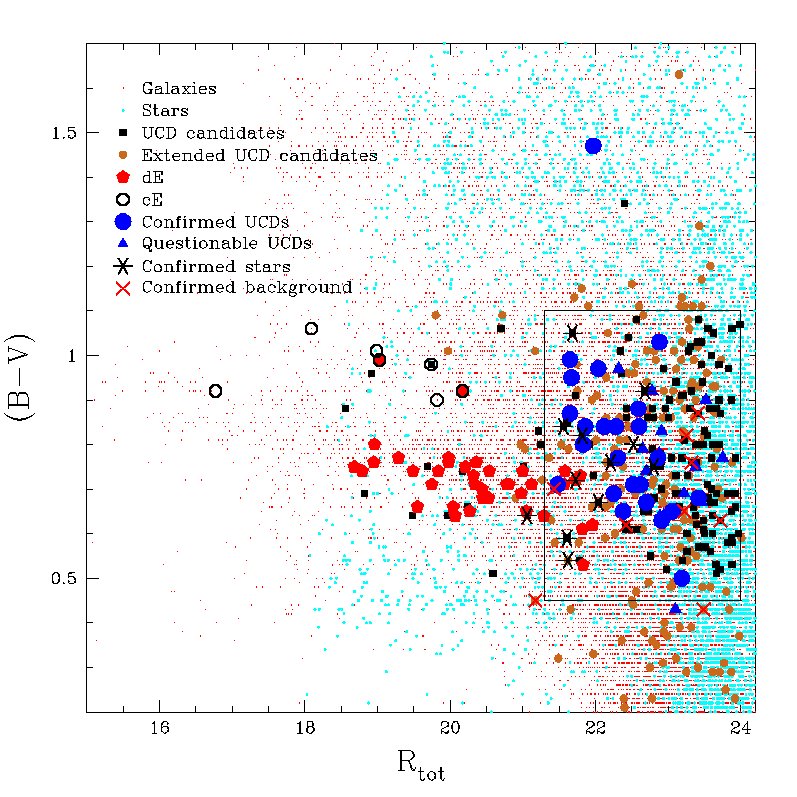}{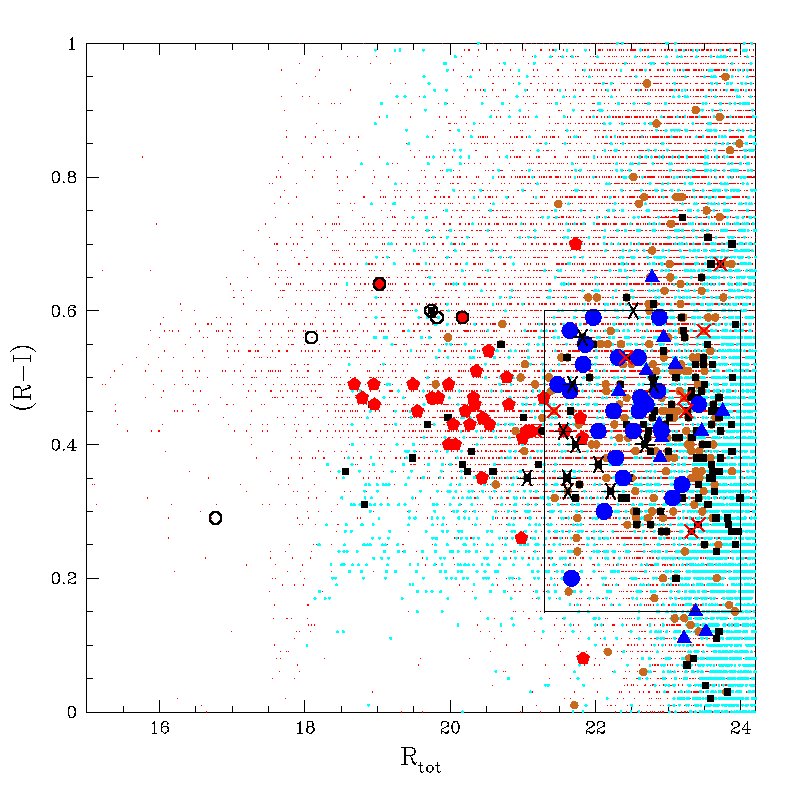}
\caption[]{Color-magnitude diagrams with UCD candidate selection criteria shown.
Photometry comes from \citet{adami}.  Objects classified as stars and
galaxies come from the full Adami et al. catalog.  The original and extended UCD
candidate samples are shown separately.   Boxes bound the
original color/magnitude cuts used to select UCD candidates within our ACS fields.  
dEs are Coma cluster members observed with the same masks as the UCDs,
while open circles represent cE member galaxies with
redshifts measured from Hectospec data \citep[Marzke et al. in prep.]{compact}.  
\label{select}}
\end{centering}
\end{figure}

In Figure \ref{select} we show color-magnitude diagrams for all
objects in our field.  Boxes delineate the original candidate color selection criteria.
Candidates, confirmed members, background galaxies, and stars are
denoted by different symbols.
dEs confirmed as
members from LRIS spectra taken concurrently with the UCD observations are also shown.
The dE red sequence is obvious in $B-V$.  UCDs meanwhile exhibit a very
large spread in colors ranging from the red sequence on the blue side 
to 0.4 magnitudes redder than the red sequence in $B-V$.  One discrepant point with very red
colors is likely affected by photometric errors.
Two cE galaxies observed during our initial LRIS campaign, along with
several others taken from \citet{compact}, also lie redward of the red sequence by about 0.2 
magnitudes.  

We find that, in good conditions, S/N starts to become too low to measure redshifts for
these high surface brightness objects at $R \sim 23.3$, 
and we fail to measure secure redshifts altogether at $R = 23.5$.
Therefore, brighter than $R = 23.3$ where we can measure redshifts for nearly 100\% of
our sources, we find that 66\% (18/27) of the original targeted candidates are 
bona fide Coma cluster members.
This is an exceptionally high success rate given that the
UCD candidates were selected primarily based only on fairly loose color criteria.
Those targeted, however, were weighted toward objects with profiles
slightly broader than pure PSFs and we believe that this boosted our success
rate.  

\begin{figure}[t]
\begin{centering}
\plottwo{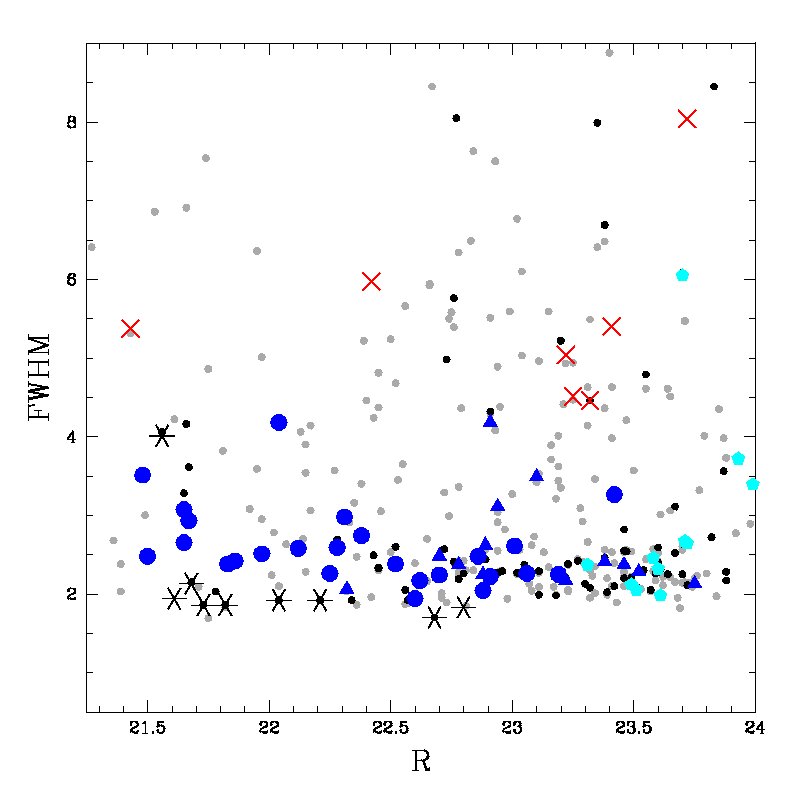}{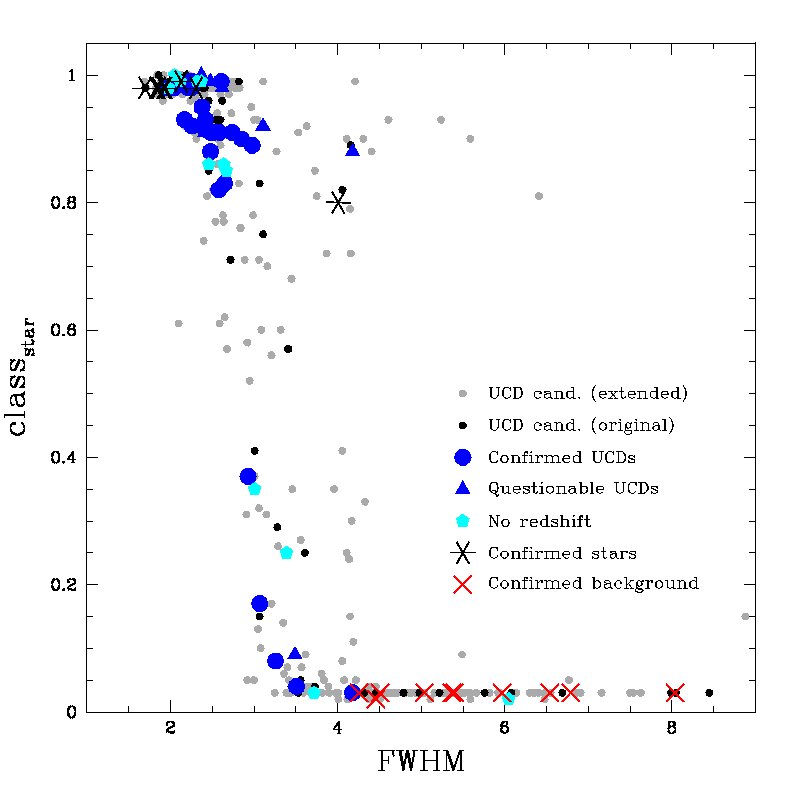}
\caption[]{SExtractor FWHM vs. magnitude (left) and SExtractor classification vs.
FWHM (right).  FWHM is in pixels, measured from our $F814W$ ACS data. 
Small pentagons are objects from our first observing run with spectra having too
low S/N to measure redshifts. 
\label{fwhmr}}
\end{centering}
\end{figure}

In Figure \ref{fwhmr}, we compare the FWHM and SExtractor classification of the UCDs as 
measured in the ACS F814W images to that of confirmed stars and galaxies. 
The ACS measurements were performed
with SExtractor as described in \citet{hammer}. 
It can be seen that the majority of confirmed UCDs
have a FWHM at a given magnitude that lies between the
stellar minimum and the larger background galaxies.  One may expect
that many of the candidates with FWHM in the range of
the confirmed background galaxies will also prove to be background
objects, while the remainder of these points with sizes larger than 
that of the stellar minimum may turn out to be further UCD cluster
members.  These partially resolved objects are classified by SExtractor in
our images as falling between stars (1) and galaxies (0).  Candidates
in this range would also be considered likely members, along with
objects classified as stars but which have slightly broader profiles.  However,
with these low number statistics, we cannot rule out the possibility
that there are large numbers of unresolved UCDs.

\subsection{Properties}\label{props}
\subsubsection{Photometric Properties}\label{propsphot}

\begin{figure}[t]
\begin{centering}
\includegraphics[angle=0,totalheight=4.5in]{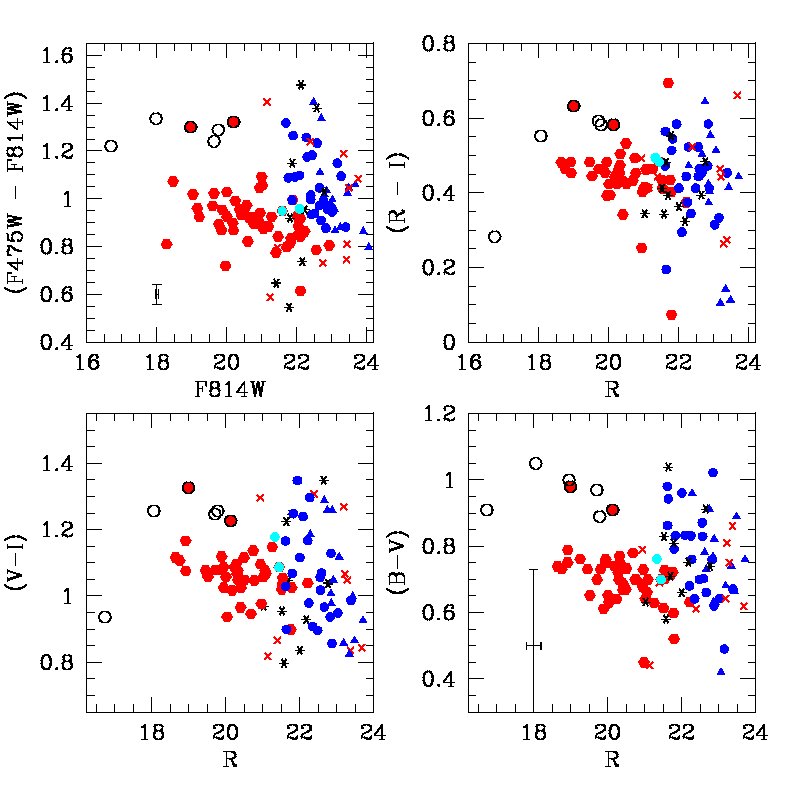}
\caption[]{Colors for all objects in our spectroscopic sample.  
$BVRI$ colors come from \citet{adami}, $F475W-F814W$ ($g - I$) from our ACS
imaging.  Magnitudes are extinction corrected using the dust maps of \citet{sfd98}.
Light circles are
objects with a faint but obvious LSB envelope around a nucleus. 
Other symbols as in Figure \ref{select}.  Representative errors for
the UCDs are provided for the ACS photometry ($g-I$ vs. $I$) and Adami et al. 
photometry ($B-V$ vs. $R$).  Photometry does
not exist for all objects in all bands, so a few UCDs are missing
from various panels in this figure.
\label{colors}}
\end{centering}
\end{figure}

In Figure \ref{colors} we show optical color-magnitude diagrams for confirmed
members of the Coma Cluster (UCDs and dEs).  In all cases,
a red sequence of normal dE galaxies, exhibiting some scatter, is apparent.  
The confirmed UCDs, which lie at the faint end of this sequence,
have a much larger spread in colors of up to 0.4 magnitudes in both
$g-I$ and $B-V$. Based on the $g-I$ ACS photometry, UCD colors range from just redward
of the red sequence to nearly 0.4 mag redder.  
Also shown in these plots are 7 confirmed Coma cluster
cEs from \citet{compact}.  These objects at brighter magnitudes also lie redward of the 
red sequence by about 0.2 magnitudes.  Although there
is a $\sim1.5$ magnitude gap between the cE and red UCD populations, these
objects do share similarly red colors and could perhaps form a single
sequence of extremely red cluster objects.  The colors may indicate
that these objects have similar stellar populations
and perhaps evolutionary histories.

\begin{figure}[t]
\begin{centering}
\plottwo{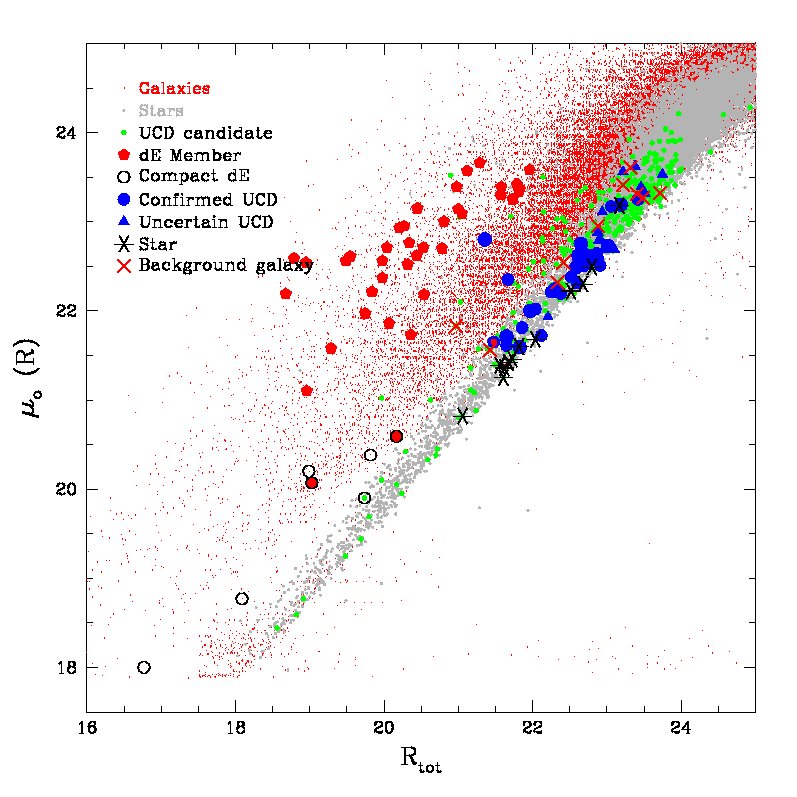}{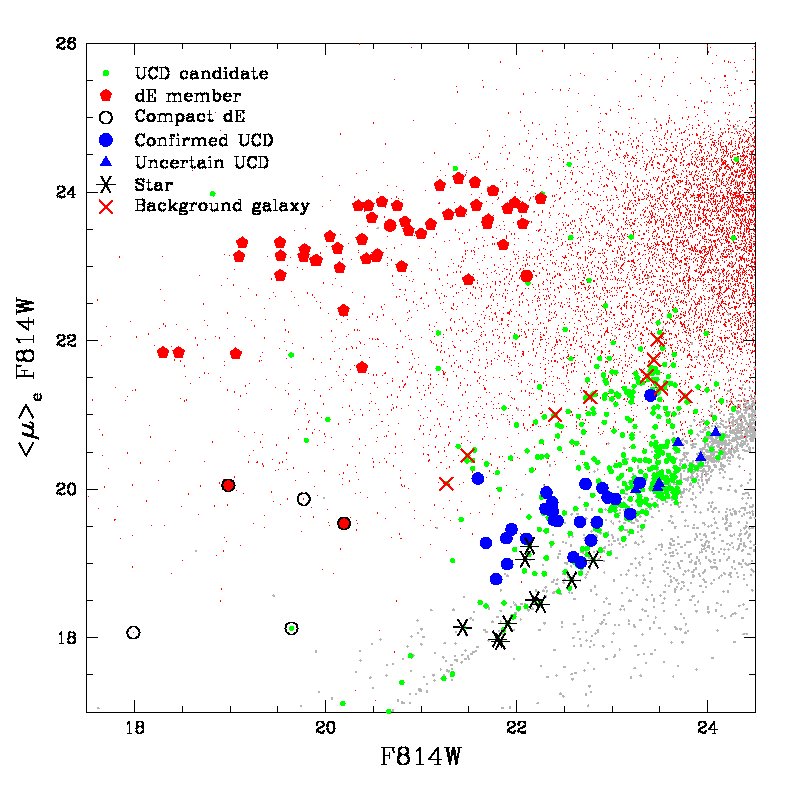}
\caption[]{Left: central surface brightness vs. $R-$band total magnitude
for all objects in our ACS survey region.  Photometry comes from
\citet{adami}. Right: mean effective surface brightness vs. $F814W-$band total magnitude
for all objects in our ACS survey region.  Photometry comes from
\citet{hammer}.
\label{gecko}}
\end{centering}
\end{figure}

We plot UCDs and other object types in magnitude-surface brightness space in 
Figure \ref{gecko}.  
Central surface brightness vs. $R-$band total magnitude comes
from ground-based photometry \citep{adami} for
all objects in our survey region while $\langle\mu\rangle_{e}$ vs.
F814W are measured from our ACS data.  LRIS confirmed normal dwarf galaxies 
are found to lie along the well known magnitude-surface brightness relation
for dE galaxies \citep{cb87,fs88,bc91,bos08}. At much higher surface brightness, 
a second almost parallel sequence of UCDs is found, bounded at high surface brightness
by the seeing disk in the ground-based data where the UCDs are merged with
the stellar sequence.  The ACS data, on the other hand, do a much better
job at separating stars and UCDs, although some overlap is still present.
In the region between the dE and UCD sequences
are confirmed and probable background galaxies (see \citet{lris}).
Brightward of the UCD sequence, the set of 7 cE galaxies 
also lie at very high surface brightness, well separated from the
normal galaxies.  Two of these objects for which we obtained LRIS spectra
are filled in.  Although the UCDs and cEs are separated
by a gap of 1.5 magnitudes, their surface
brightness characteristics also suggest that these two object types
may follow a continuous sequence and form a single class of object.

\subsubsection{Structural Properties}\label{propsstruct}

Because the UCDs are marginally resolved in the ACS data, we are able to measure the
sizes of many of these objects.  Other studies have shown that it is possible 
to measure globular cluster effective radii if the stellar PSF is accurately 
known and the cluster size (r$_h$) is larger than about 10\% of
the PSF FWHM \citep[see][and references therein]{harris10}.  The stellar
PSF for the ACS/WFC is about 0.1 arcsec, so we would expect to be able to measure
sizes down to about 5pc.  Recently, \citet{juan} identified candidate UCDs near
NGC 4874 based on size measurements of stellar-like objects in 
these ACS data.  From a comparison of measurements in two different bands,
they determined that they could accurately measure sizes down to r$_h = 9.2$ pc.

To measure the structural parameters for these UCDs, including physical 
sizes, we perform surface brightness profile fitting on the objects in our 
F814W ACS images.  
Two methods are used to fit these barely resolved 
objects. We first model each object with a Sersic function using the
two-dimensional fitting algorithm Galfit \citep{peng02}.
Galfit fits the light profile of galaxies with axisymmetric analytical functions
and is robust for well resolved,
extended objects.  Often used for profile fitting of globular clusters,
Ishape \citep{larsen} is a routine in the {\it baolab} data reduction package 
designed specifically for fitting marginally resolved sources. 
Both convolve a PSF with model profiles and
both algorithms determine best fits through $\chi^2$ minimization.

To run Galfit, the user must provide a PSF for convolution with a model,
an uncertainty map which is critical for determining when the best fit
solution is reached, the choice of profile to fit, and initial guesses for
each parameter to be fit.  The procedure we use is similar to that explained in detail in \citet{hoyos}.
Briefly, to create the uncertainty images, we make use of the Multidrizzle 
output inverse variance images and in addition take into account the
Poisson noise from the sources themselves.  
To generate PSFs appropriate for Galfit input, TinyTim \citep{tt95} PSFs are generated
for a grid of locations on the ACS chips for each filter.
These are then added at the corresponding shifted locations to blank individual 
distorted FLT 
images.  The images are then run through Multidrizzle with the same
configuration used to produce the final combined science images.   
PSF images 3 arcsec in size are extracted from this drizzle combined image.  
For the UCD fits, we perform the fitting in image
sections $300\times300$ pixels in size. All objects other than the central UCD are 
masked.  
We fit for all parameters of the Sersic function (total magnitude, effective 
radius R$_e$, index n, axis ratio b/a, and position angle) 
along with the sky value. 
We found that measured values often depended on initial guesses. 
Furthermore, the Sersic index n and R$_e$ are strongly correlated. We therefore ran
Galfit with initial guesses for the Sersic index ranging from 1.4 to 6.4.  We take the 
fit with the 
lowest $\chi^2$ value as the best fit.  Since Galfit measurement uncertainties are
in most cases underestimates,  
we take for measurement uncertainties the standard deviation of the measurement values 
from fits with 5 sets of initial guesses, including two cases where we have held the Sersic
index n fixed at 2.5 and where we included an upper limit constraint of 7.6 for the index. 

Ishape requires 10 times subsampled PSFs so we generate an empirical PSF based on 
26 real stars in one of our ACS images using the routines in the IRAF daophot package.
We fit each UCD with a set of 5 profile
shapes: a Moffat profile with power index 2.5, 
a Sersic function where the index n is fitted for, and 
King profiles with three distinct values for the concentration
parameter (defined as $r_t/r_c$ where $r_t$ is the tidal radius and $r_c$ is the
core radius): 15, 30, and 100.  We chose to fit
these as elliptical functions.   Output includes the FWHM along
both major and minor axes, position angle, and $\chi^2$ for the fit. 
To convert FWHM to effective radii, we take the circularized FWHM from the 
geometric mean of the semi-major and minor axes
and convert to R$_e$ using the appropriate concentration-dependent conversion
factors described in \citet{larsen01}, and provided in the user guide.
We take as final measurements the parameters from the fit producing the best residuals
and lowest $\chi^2$. These proved consistently to be from King models with concentration
parameters 30 and 100. During testing, it was discovered that PSF size and fitting
radius affected the size measurements by amounts greater than the small quoted
errors.  Since the quoted errors are therefore expected to be underestimates, which 
also assume a particular model is a good match to the true profile shape,
we therefore use the standard deviation of measurements from fits with 5 different models
using two different fitting radii each (10 and 25 pixels) as a more realistic measure
of the uncertainty in size.  Final sizes are taken to be those based on 25 pixel fitting 
radii which typically produced the fits with the smallest residuals.

\begin{figure}[t]
\begin{centering}
\plotone{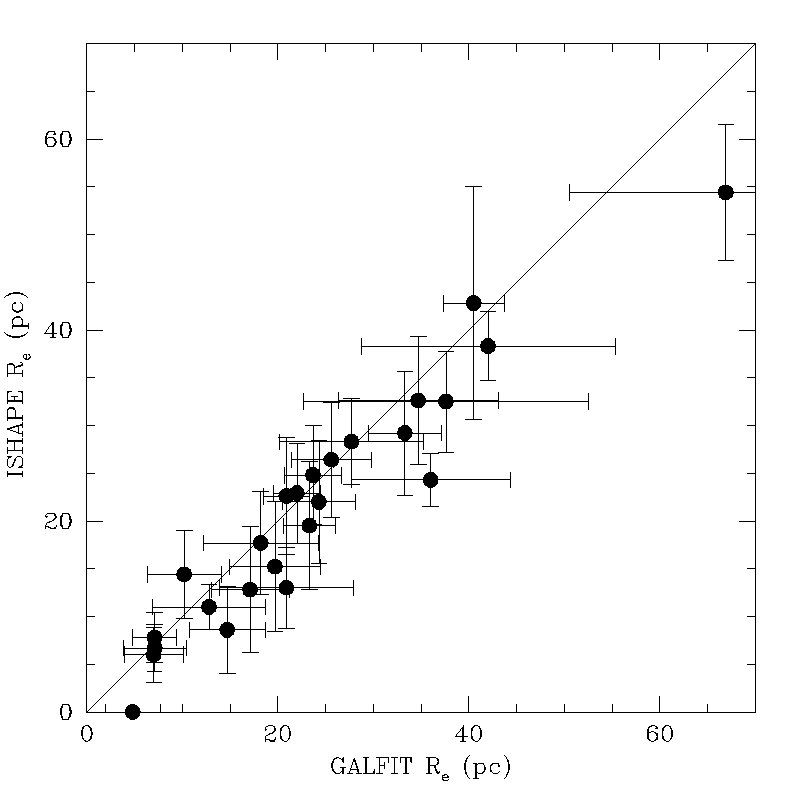}
\caption[]{Comparison of Galfit and Ishape measured R$_{e}$. One object,
151072, has a much larger measured size and is not shown here but we note that the
Galfit measurement for this object deviates significantly from that of Ishape.
\label{sizeig}}
\end{centering}
\end{figure}

\begin{figure}[t]
\begin{centering}
\includegraphics[totalheight=6in]{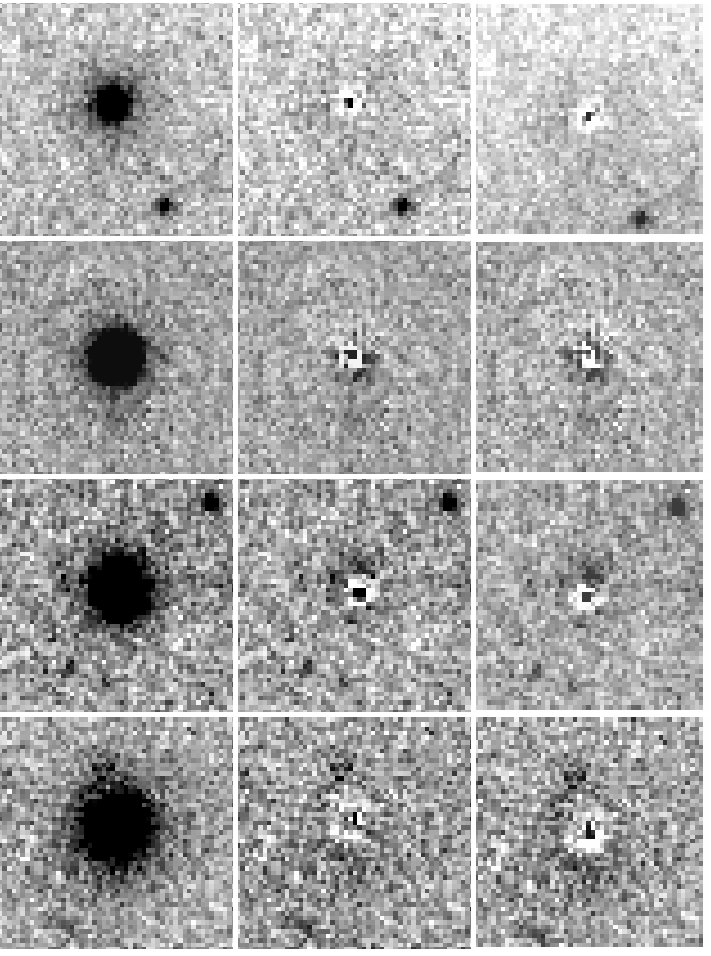}
\caption[]{Residuals from best Galfit (middle) and Ishape (right) fits for 4 UCDs.  
From top to bottom: 150000, 121666, 195526, and 151072.  Panels are 2.5 arcsec across.
Strong residuals are apparent and may indicate that two-component fits are 
warranted in several of these cases.
\label{resid}}
\end{centering}
\end{figure}

\begin{figure}[t]
\begin{centering}
\includegraphics[totalheight=6in]{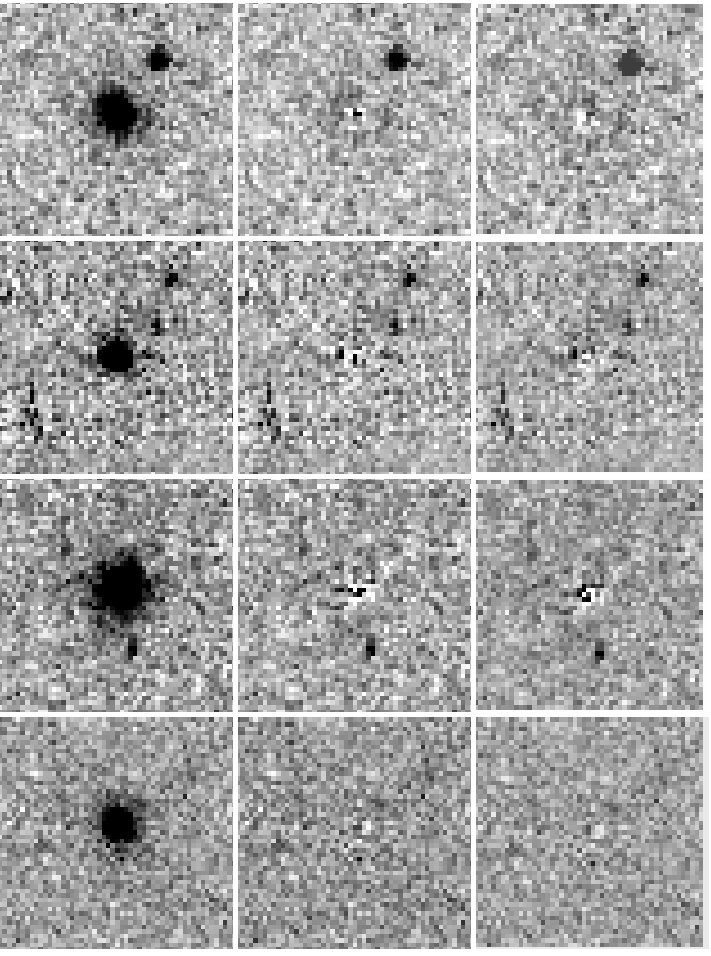}
\caption[]{Residuals from best Galfit (middle) and Ishape (right) fits for another 4 UCDs.  
From top to bottom: 1041508, 182204,195614, and 163400.
Panels are 2.5 arcsec across.
\label{resid2}}
\end{centering}
\end{figure}

Size measurements are presented in Table \ref{ucdsize}.  
To convert effective sizes from arcsec
to parsec we assume a distance of 100 Mpc.  We find sizes ranging from
$5 - 125$ pc with a median effective radius (for both Galfit and Ishape measurements) of 23 pc. 
Overall, the consistency between
Galfit and Ishape measurements of R$_e$ is quite good.  We show the comparison
in Figure \ref{sizeig}.  For the largest object, 151072, Galfit finds a size about
twice as large as Ishape.  
This may be an indication that the object has an extended halo which affects the Sersic
function fits more than the King profile fits.  Although we only fit single component
model profiles for these objects we note that a number of the objects, including some
of the larger UCDs, may be better fit with two components.  For objects best fit
with very large n, this may be caused by a lower surface brightness envelope 
surrounding a high surface brightness core forcing a fit with larger wings.
In particular, object 150000, although highly obscured in the diffuse light of a nearby
bright galaxy, upon closer inspection appears to have a large low surface brightness envelope.
Objects 121666 and 195526 also may require a second component fit. 
In Figure \ref{resid}, we show the residuals from both Galfit and Ishape fits
for 4 UCDs which may be better fit with two components.  Object 150000
in the top panel has 19\% of the flux remaining in the residual when fit with
a single Sersic function. UCD 151072 in the bottom panel has residuals at the level
of 5\% when fit with a King profile. Figure \ref{resid2} 
displays residuals for 4 cases where the UCDs are well fit with single component 
models by both Galfit and Ishape.

Three of the objects less well fit by single component profiles turn out to be the
three brightest UCDs in the sample. If the poor fits are due to an extra envelope of 
material surrounding the UCD core, we could be seeing remnant tidal debris from
the stripping of nucleated, more massive galaxies.  Extended halos have also been
found around UCDs in Virgo and Fornax from HST imaging \citep{Ev08} and around massive
GCs in the Milky Way and other nearby galaxies \citep{mvm05,mcl08}. They may therefore be 
a common property of star clusters
and are not necessarily evidence for the threshing mechanism.  In fact, the lack of 
such signatures around most of the fainter UCDs in our sample may simply be due to 
the lower S/N and the small sizes of these objects.  If the brightness and 
extent of the envelope is correlated with UCD size and brightness, extended halos around
fainter objects could remain undetectable in our imaging.  Object 150000, on the other
hand, has an intermediate core size and brightness but more extended low surface brightness
envelope, distinguishing it from the rest of the objects in the UCD sample.

The Sersic index n measurements listed in Table \ref{ucdsize} range from 1.7 - 7.6
with the majority having n $> 4$.  Since typically only giant elliptical galaxies
are best fit with such large n, and since R$_e$ and n are known to be correlated,
we compare those results to fits with forced n $= 2$, more typical of globular clusters
in the Milky Way \citep{mcl08}.  For most UCDs the resultant R$_e$ was little changed.  Excluding
4 cases, the average difference in R$_e$ between the best fit listed in the table
and a fit with forced n $= 2$ is only 0.1 pc.  This suggests that for such small
objects, the shape parameter has little effect on the fit, and cannot be well constrained.
For 4 objects, the measured size with n $= 2$ was significantly smaller (163400, 
151072, 150000, and 1043225 with sizes 41.1 pc, 67.6 pc, 22.7 pc, and 11.9 pc, respectively), 
more in line with the Ishape size measurement.  In at least 2 of these cases, the larger  
R$_e$ (and Sersic index) from the unconstrained fit is likely due to the presence of an outer 
envelope.

\subsubsection{Ages and Metallicities}\label{propsmetal}

From our spectra, we measure absorption line index strengths in the Lick/IDS
system and compare to SSP models to derive luminosity weighted metallicity
and age estimates for our confirmed UCDs. We have chosen to use the
models of \citet{schiavon} and the publicly available EZ-Ages code \citep{graves} 
for deriving the ages and abundances.  An advantage of EZ-Ages is that a 
separate code within this package measures line indices from our spectra. 
These models also include the effects of non-solar abundance patterns.  

Line index strengths from our LRIS blue-side spectra are measured after first degrading 
the spectra to the Lick resolution.  These line strengths are fed into the EZ-Ages 
code which first determines an initial estimate for the age and metallicity 
from the H$\beta$ and Fe indices using grid inversion, and then calculates
the [Mg/Fe] ratio to obtain an alpha element abundance measurement.  Age and 
metallicity are re-derived from H$\beta$ vs. Mgb, where the Mgb index is highly 
sensitive to alpha element abundance.  If significantly different from the first 
age and metallicity measurements, the [Mg/Fe] ratio is increased incrementally 
and iterations proceed until a user supplied tolerance is
reached. The process is then repeated for other Lick indices. 
The models span a wide range of age and metallicities, but will fail to produce derived
quantities in cases where measured indices fall outside of the model range. 

As initial input for the fitting process, we use a Salpeter IMF
and solar scaled isochrones.  Velocity dispersions
are also required but due to the low resolution of our spectra, we have not attempted to
measure these for our UCDs.  However, velocity dispersions have been measured for a number of 
UCDs in the Virgo and Fornax clusters \citep{has05,mieskeb08,Ev07a,hil07} and found to range
between $9 < \sigma < 42$ km/s.   We therefore simply assume a small
$\sigma_v$ of 20 km/s.  We confirm that including velocity dispersions up to 65 km/s 
(the mean value found by \citet{compact} for 7 cE galaxies and presumably a very high
upper limit for our much smaller UCDs) affects derived values insignificantly within 
the uncertainties provided by EZ-Ages.
As we did not obtain any Lick standards we cannot correct for any systematic offsets.  
Although the $\langle$Fe$\rangle$ index (an average
of Fe $5270\AA$ and Fe $5335\AA$ measurements) is often used to estimate 
metallicity, the Fe $5335\AA$ line 
is near the edge of our blue-side spectra or missing altogether.  We therefore 
make use only of the Fe $5270\AA$ line index.  The strong Mgb line is used 
for the Mg measurements. 

\begin{figure}[t]
\begin{centering}
\includegraphics[totalheight=5in]{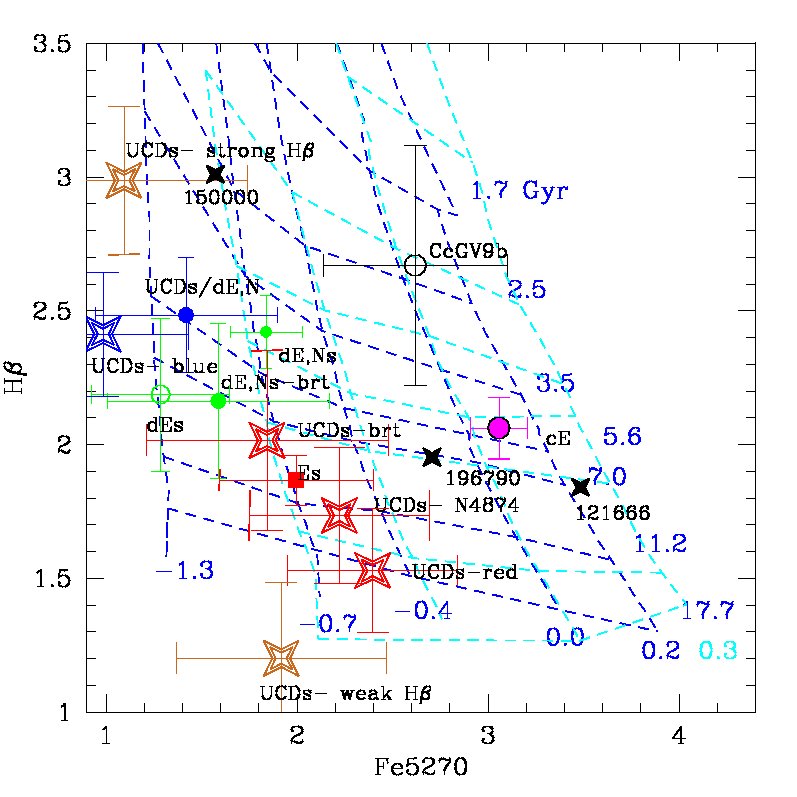}
\caption[]{Measured Fe $5270\AA$ vs H$\beta$ line strengths.
[Fe/H]-age grid models are from \citet{schiavon} 
for $\alpha/Fe = 0$ (blue) and 0.3 (cyan).  
Plotted are Fe 5270 and H$\beta$ measurements for composite spectra
of similar object types.  Object types are labeled on the plot.  We include two points for
dE,N: the larger solid circle includes 4 objects with bright, prominent nuclei, the smaller
one includes 24 dE,N with small, faint nuclei.  The open black circle comes from 
\citet{compact} for the same object labeled as cE below it.   Star-like symbols 
refer to UCDs with different combinations of stacked spectra:
the 5 brightest UCDs, 6 red ($V-I > 1.05$),
9 blue ($V-I < 1.05$), 5 red UCDs around NGC 4874, 8 with weak H$\beta (< 2.5)$, and
8 with strong H$\beta (> 2.5)$. The UCD/dE,N includes object 151072 from this paper
and 242439 from \citet{lris}.  Both are compact sources with a hint of an extended 
envelope surrounding them. We also plot 3 individual bright UCDs (small stars) but do 
not include the large error bars.   
\label{metals}}
\end{centering}
\end{figure}

We initially ran EZ-Ages on our UCDs individually, but as most have
S/N $< 15$, we found results unreliable, having exceptionally large 
uncertainties for the derived ages and metallicities.  We therefore 
stack multiple UCD spectra.  For this to be useful, the objects must have similar stellar
populations and spectral properties.  Otherwise, this would only produce a spectrum with
a random mix of line strengths. Therefore, we stack
spectra of similar type objects based on physical properties such as color
and location in cluster, and on properties of the individual line index measurements.
Different sets of UCD spectra are combined.  We have stacked the spectra from 
the 5 brightest UCDs, 6 red ($V-I > 1.05$) UCDs, 9 blue ($V-I < 1.05$) UCDs, 5 red UCDs 
around NGC 4874, 8 with weak H$\beta (< 2.5)$, and 8 with strong H$\beta (> 2.5)$.
Several UCDs do not have $V-I$ measurements and are not included in the color selected
samples, and we do not include any UCDs from our second observing run as the 
S/N was particularly low.  We plot H$\beta$ and Fe $5270\AA$ line index measurements 
from the composite spectra in Figure \ref{metals}.  
We also include line index measurements for 3 individual UCDs with high S/N spectra which 
suggest population ages or abundances that are different
from the majority of the UCDs in our sample.  The
uncertainties for these are large, however, with error bars which span much of the grid.  
For clarity, we therefore do not include error bars for these individual objects.
Overlaid in the figure are grids for $\alpha/Fe = 0$ and 0.3 with lines of constant age
and [Fe/H].

Spectra for non-UCDs come from \citet{lris}. Included in this figure are measurements for
elliptical, dE, and dE,N galaxies.   To produce these, we have stacked spectra for
2 small elliptical galaxies, 10 low surface brightness dE galaxies, 4 dE,N with 
prominent nuclei and 24 dE,N with small/faint nuclear star clusters.
For the 4 objects with prominent nuclei, we re-extracted
the two dimensional spectra with smaller apertures meant to include light primarily
from the nucleus although some contamination from the spheroidal component of these dE,N may
still be present.  We have also measured Fe $5270\AA$ and H$\beta$ for one of two
cEs for which we have obtained spectra (object 91543 in \citet{lris}).  This object was also
observed with Hectospec on the MMT and line index measurements are presented 
in \citet{compact}.  Both measurements are included in this plot.  

Table \ref{Tmetals} presents line index measurements and derived
ages and metallicities ([Fe/H]), along with [Mg/Fe], an indicator of alpha element
abundance.  For comparison, we include in this table [Fe/H] estimates from the $V - I$ colors
using an empirical relation based on Galactic globular clusters from \citet{barmby00}.
For the composite spectra cases, we use an average $V - I $ color.

Given the large uncertainties, it is clear that our age and metallicity measurements should not
be taken at face value.  Furthermore, we find that the addition of single UCDs in
composite spectra in some cases produce a large change in measured values.
However, we can draw some conclusions from this exercise.  The most obvious is that the
UCDs are not a homogeneous population.  Rather, they exhibit a spread in age and metallicity
significant at the $\geq 2.5\sigma$ level.   This was already expected based on the large 
color spread we find for these UCDs.
Line index measurements confirm this range is due to real differences in stellar 
populations.  Redder UCDs are found to be both older and more metal rich than
the bluest UCDs.   Blue UCDs have on average intermediate ages and low metallicity.
The majority have sub-solar 
metallicity, although a couple individual cases show evidence for more metal 
rich stellar populations.  Red UCDs around NGC 4874 are found to be no different
from the overall set of red UCDs.  

With the exception of a subset of red UCDs, most UCDs appear to have high
[Mg/Fe] abundances, suggesting possible super-solar alpha element abundances. This 
would indicate that the stellar populations in UCDs formed in quick bursts of star
formation.  Because of the large uncertainties for this derived quantity, we caution the 
reader against drawing any firm conclusions.  However, we note with interest that
the derived values for most subsets of UCDs are similar to those of globular clusters
which typically have super-solar alpha element abundances. See e.g. 
\citet{puzia05} who find a mean [$\alpha$/Fe] $= 0.47\pm0.06$ with a dispersion of 0.26 dex
for globular clusters in early type galaxies. 
The set of strong H$\beta$ UCDs are found to be very young, metal poor, and highly
alpha element enriched.  The relation between star formation time scale and 
alpha element abundance in \citet{thomas05} would suggest a very 
rapid burst of star formation with a timescale of only $\sim0.25$ Myr.  
The extreme values for this set of objects is likely strongly influenced by the inclusion
of object ID 150000.  This object appears to be surrounded by a very faint extended envelope 
and may not be the same class of object as these other UCDs.

Comparing dE to dE,N derived ages and metallicities, we do not find any differences 
within the uncertainties.  
Red UCDs are generally older than the dE and dE,N galaxies, significant at 
greater than $2\sigma$.  There is a slight 
hint that the UCDs are more metal rich as well, but this is
insignificant within the errors. Blue UCDs are similar to the dE and dE,N within the errors, 
having slightly lower metallicities and similar ages.  
The measured indices we find for the one cE are quite different from
the UCDs, having super-solar metallicity and a young to intermediate age.
Other cEs from \citet{compact} have metallicities as low as those found for the red
UCDs.

\subsection{Distribution}\label{distro}

Looking at the spatial distribution displayed
in Figure \ref{overlay}, we find that all confirmed
UCDs lie toward the central core region of the cluster along a
band in declination at $\delta \sim 27.97$.
We do not find any members lying far from this, although it can
be seen that very few UCD candidates
were targeted away from this band.  Three crosses (denoting
background galaxies or foreground stars) lie north
of 28.06 and another three south of 27.94.  With these small number statistics
it is hard to say too much about the true spatial distribution of the full
UCD population.
However, if we look at the original set of good candidates (from which we achieved
a 66\% confirmation success rate) in conjunction with the confirmed
members, we find a pronounced linear structure
running E-W from NGC 4874, past the other central giant elliptical NGC 4889, and continuing
toward IC 4051.  For a comparison, we take a sample of 10,000 points
distributed randomly over the fields which were available during the initial
selection of candidates, assuming a uniform distribution.  A 2-dimensional
Kolmogorov-Smirnov (KS) test finds that the good candidates have a distribution
different from a uniformly distributed sample at a 99.95\%
confidence while a 1-dimensional KS test finds that the candidate population differs
in declination at 99.997\% confidence.  In contrast, the candidates have a different
distribution in RA than the random sample at only 67.18\% confidence.
As it is hard to imagine how the color selection criteria
could have biased the candidate list spatially, we suspect this is a
real structure in the Coma cluster core.  It is possible, however, that this
apparent linear structure is produced primarily by UCDs associated with giant
galaxies in the cluster core.

We also see what appears to
be a concentration of confirmed UCDs around the cD galaxy NGC 4874.
A majority 7 of 9 objects around NGC 4874 have
red colors ($V - I > 1.05$), while UCDs east of this grouping have a greater percentage of
blue $V - I$ colors (at least 10 of 16 objects).
We also find that all UCDs near NGC 4874
have $v_r > 6800$ km s$^{-1}$.  Objects eastward of this display a larger
range in measured radial velocities.  

\begin{figure}[t]
\begin{centering}
\includegraphics[totalheight=2.in]{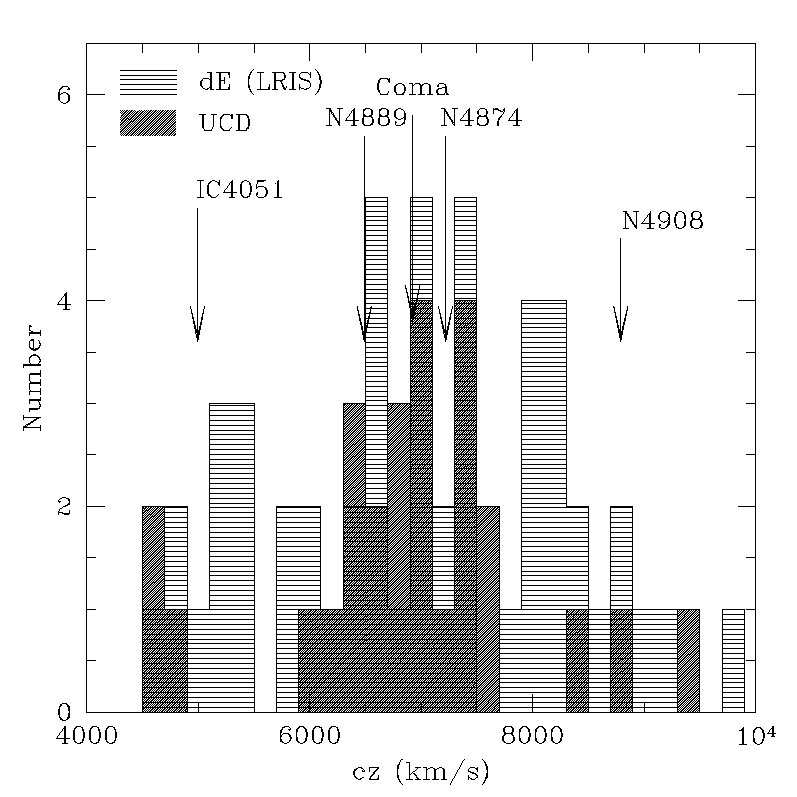}
\includegraphics[totalheight=2.in]{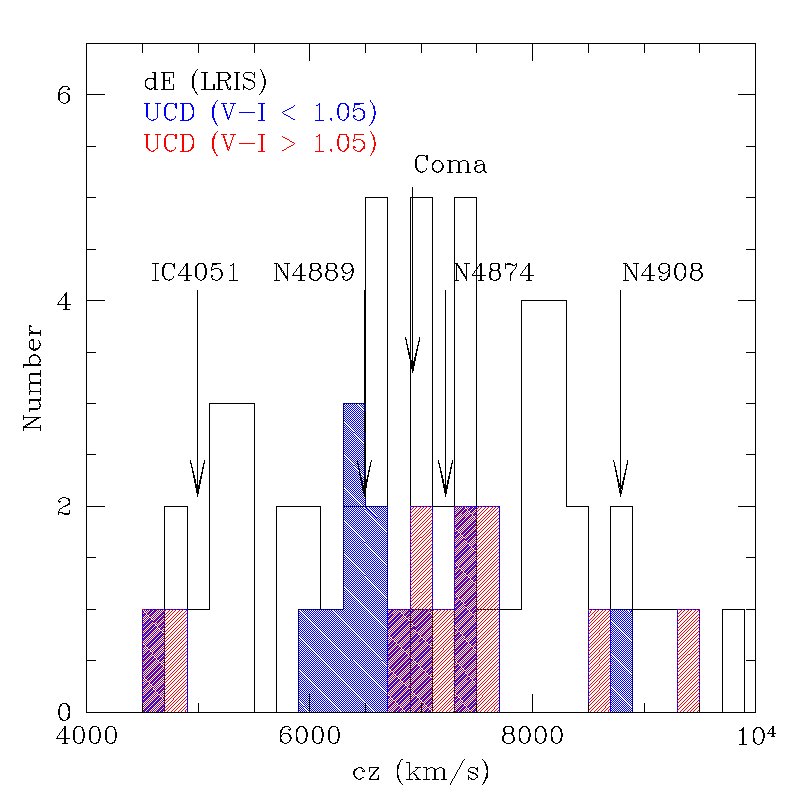}
\includegraphics[totalheight=2.in]{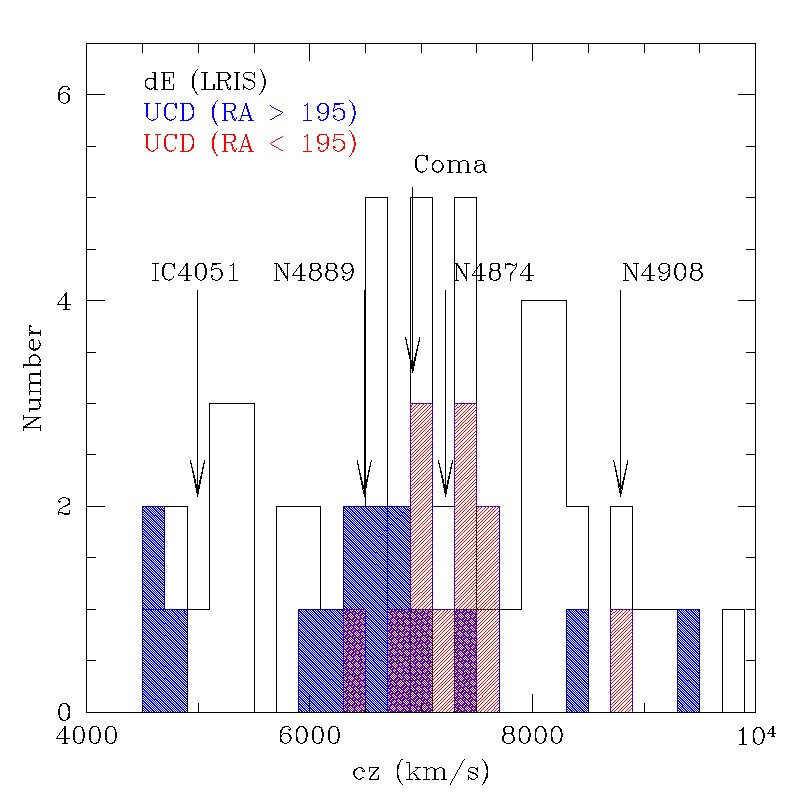}
\caption[]{Histograms of UCD radial velocities. 
Left: UCD radial velocity distribution compared to that of dEs with membership
spectroscopically determined from the same LRIS MOS observations.  The location in 
velocity space of several prominent giant galaxies are labeled.
Middle: UCDs separated into blue and red populations by $V-I$ color.
Right: UCDs split by RA.
\label{prophist}}
\end{centering}
\end{figure}

Turning to the radial velocity distribution, we note that line-of-sight velocity 
dispersions for UCD populations in other clusters
have been found to be smaller than for other cluster galaxy types \citep{mieske07,gregg09,firth08},
indicative of strong clustering.
In the Coma cluster, \citet{ecbcmp02} have measured velocity dispersions
for the giant and dwarf galaxy populations of $979\pm30$ and $1096\pm45$ 
km s$^{-1}$, respectively.   The Coma cluster has a mean radial velocity of 6925 km s$^{-1}$. 
For the faint, low surface brightness sample of 51 dwarf cluster members 
observed with LRIS, we find
$\langle v_r\rangle = 6970\pm 178$ with $\sigma_v = 1269\pm126$ km s$^{-1}$ \citep{lris},
slightly higher than previous measurements. 
For the UCD population, we find a mean radial velocity of 
$\langle v_r \rangle = 6887\pm207$ and a velocity dispersion
$\sigma_v = 1072\pm146$ km s$^{-1}$, comparable to what had been found for
giant and brighter dwarf galaxies, and smaller (although by $< 2\sigma$) from what we 
find for dE and dE,N galaxies in the same region. 
Histograms showing the radial velocity distribution of the UCDs along with
that of the normal dwarfs observed in the same masks are presented in 
Figure \ref{prophist}.  Peculiar velocities of prominent galaxies are indicated.
The left histogram shows that most UCDs have velocities similar to, and centered on,
the cluster mean, with just a few outliers. 

The isolation in color, space, and velocity noted in the spatial distribution is also 
evident in the radial velocity histograms
(Figure \ref{prophist}) where there appears to be a distinction between
the UCD groups split either by color or RA with hints of associations with
specific major galaxies.  
For example, 12 UCDs with RA $< 195$ have $\langle v_r\rangle = 7296\pm160$ km s$^{-1}$
with $\sigma_v = 558\pm119$ as compared to 15 objects at RA $> 195$ with
$\langle v_r\rangle = 6559\pm325$ km s$^{-1}$ and $\sigma_v = 1257\pm238$, a difference 
of about $3\sigma$ in the velocity dispersion.   Table \ref{ucdvels} lists
mean radial velocities and dispersions for different UCD subsamples.

\begin{figure}[t]
\begin{centering}
\includegraphics[totalheight=2.5in]{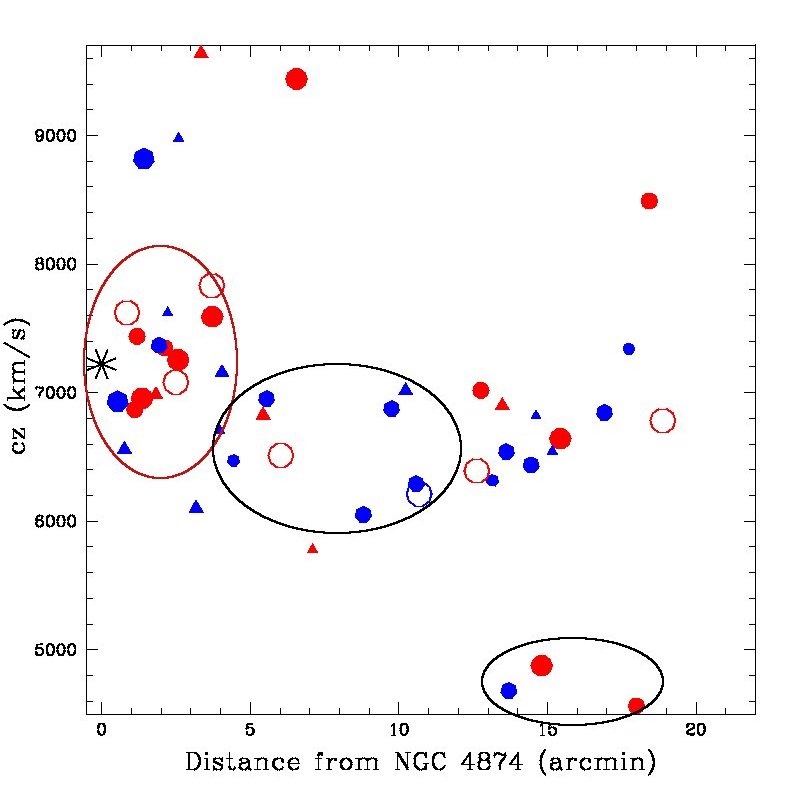}
\includegraphics[totalheight=2.5in]{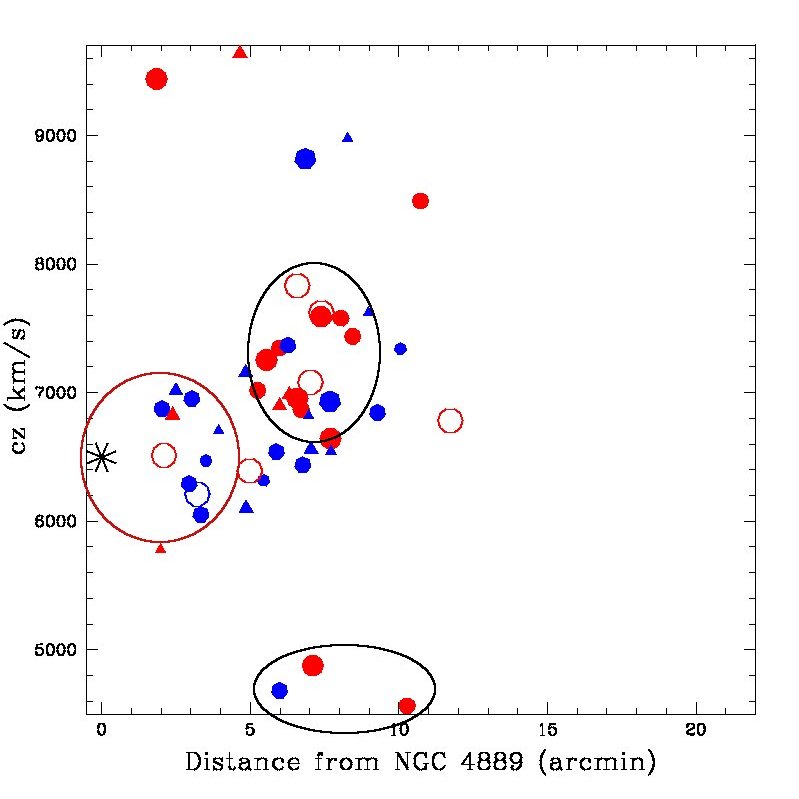}
\includegraphics[totalheight=2.5in]{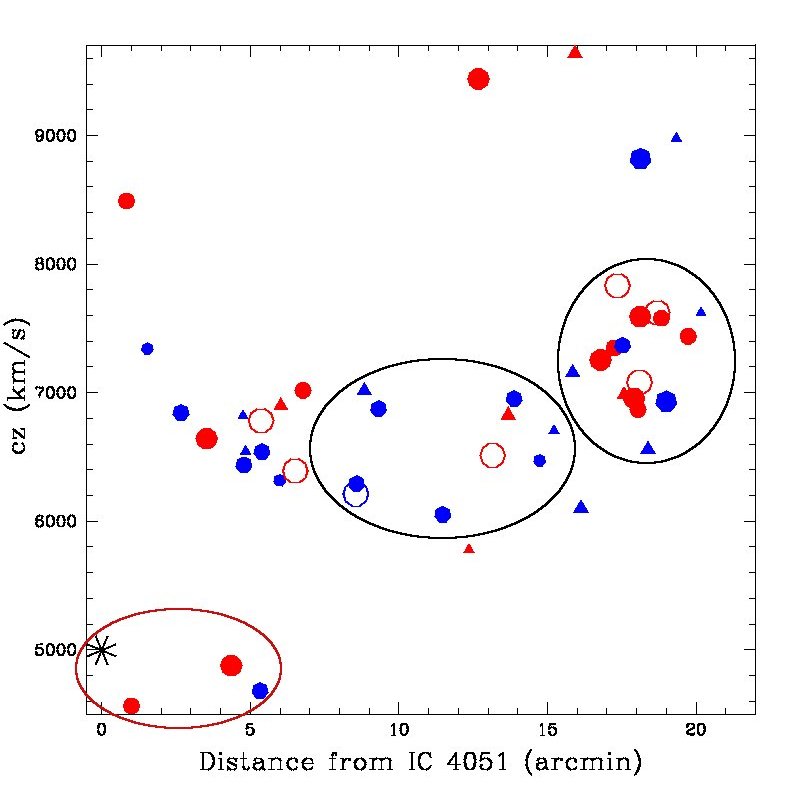}
\includegraphics[totalheight=2.5in]{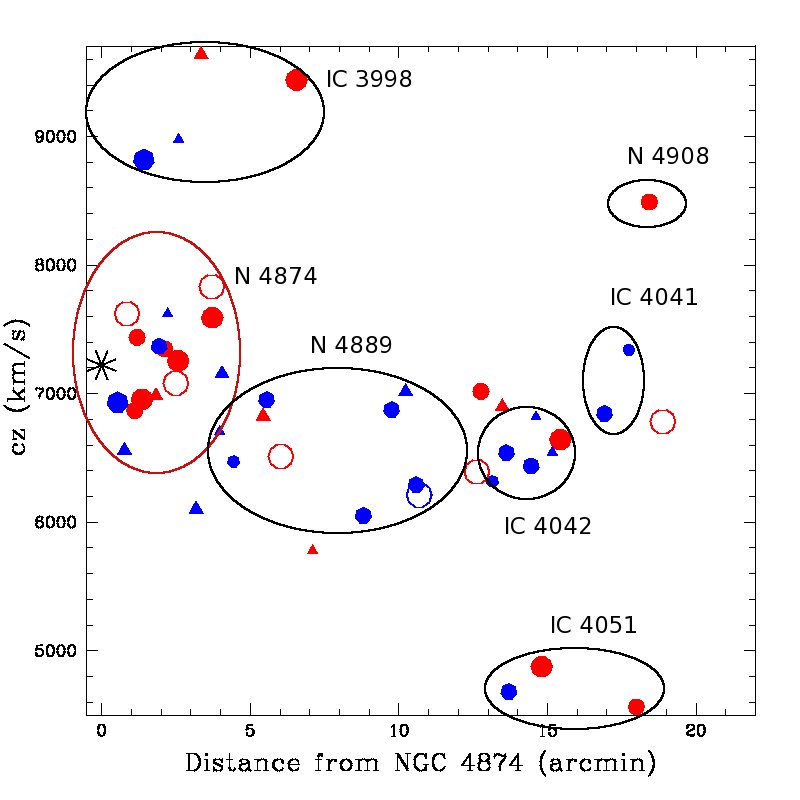}
\caption[]{UCD radial velocity as a function of projected distance from prominent Coma
cluster giants (asterisks) including the cD galaxies 
NGC 4874 and NGC 4889, and IC 4051 (a galaxy with one 
of the highest Coma cluster globular cluster specific frequencies). 
One arcmin corresponds to 29 kpc.  
Red symbols have $(V-I) > 1.05$, while blue have $(V-I) < 1.05$. Where $V-I$ colors do
not exist, we use colors in other bands to infer a rough $(V-I)$ color. Larger
symbols represent brighter magnitudes.  Open circles are cEs, triangles are 
objects with insecure redshifts.  Objects which may be associated with
individual galaxies are encircled.  The bottom right panel displays all possible associations
of UCDs with giant galaxies.
\label{dstn}}
\end{centering}
\end{figure}

Putting these distribution trends together in Figure \ref{dstn}, we show UCD projected 
distance from 3 prominent Coma giants vs. radial velocity and
see evidence that many of the UCDs are associated with giant
ellipticals.  In particular  
9 UCDs in velocity and spatial proximity to NGC 4874 have $\langle v_r\rangle = 7257\pm87$ 
with $\sigma_v = 254\pm60$ km s$^{-1}$. This
is very similar to the radial velocity, $v_r =  7220$ km s$^{-1}$, of NGC 4874 
and over $3\sigma$ from the cluster mean,
indicating these objects are more likely to be associated with the 
cD galaxy NGC 4874 than the general cluster potential.  A study of Coma cluster
globular clusters by \citet{peng09} finds that the GC population of NGC 4874 extends
out to $\sim130$ kpc before the intracluster GC population starts to dominate.  This 
corresponds to about 4.5 arcmin.

We calculate whether the 9 UCDs near NGC 4874 are likely to be bound to the cD galaxy. 
The mass of NGC 4874 has been estimated at 
$1.4 \times 10^{13}$ M$_{\odot}$ from X-ray
observations \citep{vfj94}.  We take $1.0 \times 10^{13}$ M$_{\odot}$ as a lower
limit.  The escape velocity for NGC 4874 is then given by 
\begin{equation}
v_{esc} = \sqrt(2GM / r_{sep}).
\end{equation}
For the UCD at the largest separation from NGC 4874 (121666, 3.7 arcmin),  we
find $v_{esc} = 863$ km s$^{-1}$ at a projected distance of 108 kpc.
The difference in radial velocities for the 2 objects is 369 km s$^{-1}$,
lower than the escape velocity and therefore consistent with being a bound satellite. 
As in \citet{firth07}, we confirm that this object also lies within the tidal radius 
of NGC 4874 as imposed by the other central giant galaxy NGC 4889.
\citet{vfj94} find similar masses for NGC 4889 and NGC 4874.  From 
\begin{equation}
r_t = (m/3M)^{1/3} D 
\end{equation}
\citep{bt87} where $r_t$ is the tidal radius, m and M are the masses of NGC 4874 and
NGC 4889, respectively, and D is the separation, we find
a tidal radius of $\sim 5$ arcmin.   All 9 UCDs are within this projected distance,
and with a velocity dispersion of only 254 km s$^{-1}$, we expect these UCDs to be bound to NGC 4874. 

Although clustering in both velocity and spatially around NGC 4889 is not
as clear, the 5 UCDs in closest proximity, within 4.5 arcmin of NGC 4889 at 
$v_r =  6495$ km s$^{-1}$, have $\langle v_r\rangle = 6526\pm148$ km s$^{-1}$ 
with $\sigma_v = 332\pm105$.  This is also nearly  $3\sigma$ from the cluster mean.  
The lack of confirmed UCDs within 2 arcmin of this giant elliptical is strictly 
due to the fact that the ACS field centered on this galaxy was never observed. 

Three very low velocity UCDs which are spatially coincident with 
IC 4051 have $\langle v_r\rangle = 4706\pm66$ km s$^{-1}$ with $\sigma_v = 114\pm47$, 
similar to the very low peculiar velocity of 4779 km s$^{-1}$ of this giant.  This 
galaxy, which is not a central dominant giant, is remarkable for having one of the highest globular
cluster specific frequencies (S$_N = 12.7$) in the Coma cluster \citep{mfaa01}.  Two of
the UCDs have a fairly large projected separation from this galaxy.  It is possible that 
while these UCDs may not be bound to this giant, they may be part of a cluster 
sub-structure which includes IC 4051.

In addition, two objects with particularly large radial velocities are found within 4 arcmin
of IC 3998 having a similarly large radial velocity.  We also identify several
UCDs which are coincident with the galaxies IC 4042 and IC 4041 toward the eastern side
of the cluster.  As the systemic velocities for these two galaxies are similar 
to that of the cluster mean, it is possible that these UCDs simply 
belong to the general cluster potential.  
Searching for all possible associations of UCDs with giant galaxies, we find nearly all UCDs
have potential host giant galaxies (Figure \ref{dstn}).

\begin{figure}[t]
\begin{centering}
\includegraphics[totalheight=3.in]{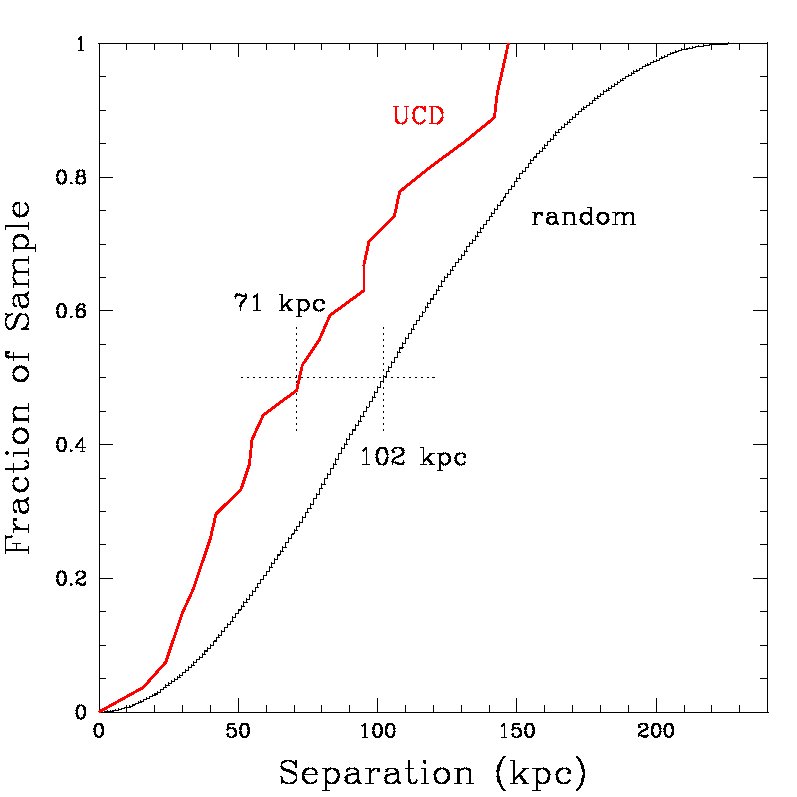}
\includegraphics[totalheight=3.in]{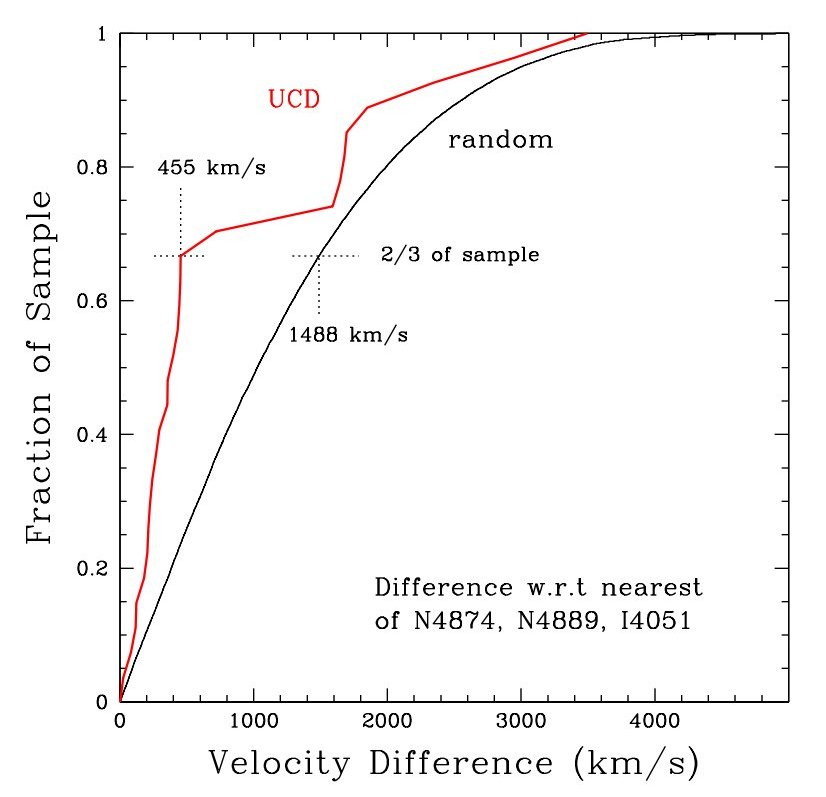}
\caption[]{Left: Cumulative distribution of the separation of each
UCD from the nearest of one of 3 giants in the core region: NGC 4874,
NGC 4889, and IC 4051.  25,000 points randomly distributed  
within the LRIS footprint are generated to produce the expectation for
a uniform distribution.  Right:  Cumulative distribution of
the velocity difference between each UCD and the closest of the 3 giants.
The curve labeled 'random' displays the same for 25,000 points with
randomly sampled velocities assuming a Gaussian distribution with mean 
6925 km/s and dispersion 1000 km/s.
\label{cumdist}}
\end{centering}
\end{figure}

In the left panel of Figure \ref{cumdist}, we show the cumulative distribution of confirmed
UCDs as a function of distance from the nearest of one of the 3 brightest galaxies in the 
core region (NGC 4874, NGC 4889, and IC 4051).   For comparison we determine the expectation 
for a uniform distribution.  This is done by randomly distributing 25,000 points over the
LRIS footprint and normalizing the resultant minimum distance  
cumulative distribution.  We find that the UCDs are more concentrated around the giant
galaxies than a spatially uniform distribution would predict.  The right panel is similar
but includes velocity information.  In this case, the cumulative velocity difference from
the nearest of one of the 3 giants is shown.  Velocities are attached to the random sample of
25,000 points by assuming a Gaussian distribution for the velocities 
having $\langle v_r\rangle = 6925$ km/s and $\sigma = 1000$ km/s.  Again we find
that the UCDs are more strongly clustered around the giant galaxies.  A clear 2/3 of the
confirmed UCDs are associated with one of 3 central giants.

With the exception of the 2 central cD/giant ellipticals, other UCDs may not be bound to 
their respective neighboring galaxies.  However, if this is the case, the velocity
and location groupings we find suggest that either these are previously bound
clusters in the process of dissolution to the cluster potential, or that 
they belong to substructures present in the Coma cluster.  

\section{Discussion}\label{disc}

From an initial observing run we identified 19 members selected primarily based on color
along with some information on profile size.  For targets with $R < 23.3$, our completeness 
limit, we had an unexpectedly high 66\% success rate for these initial observations.  
A follow-up run with an expected lower confirmation rate due to intentionally loose 
selection criteria confirmed a further 8 UCDs from poor weather observations.
These results suggest that the Coma cluster
harbors a large population of UCDs, at least in the core region, 
although this high success rate must be due at least in part
to the fact that we can marginally resolve these objects
with the superb resolving capabilities of HST/ACS.

\begin{figure}[t]
\begin{centering}
\includegraphics[angle=270,totalheight=4.0in]{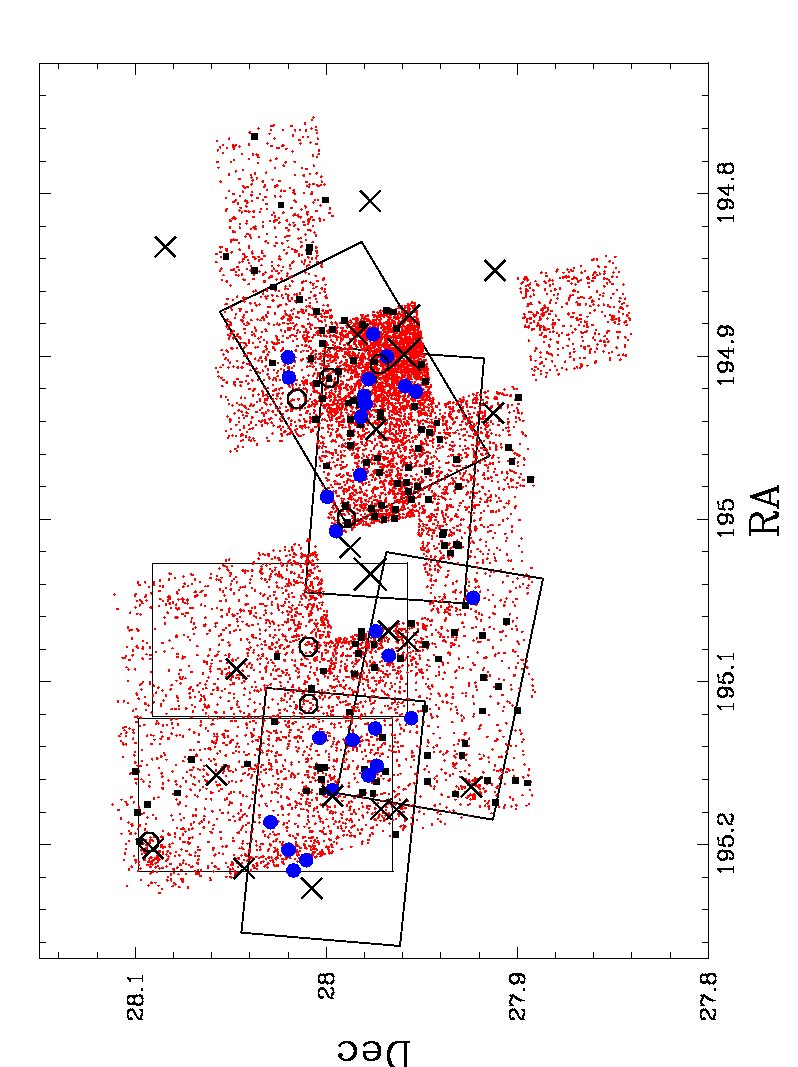}
\includegraphics[angle=270,totalheight=4.0in]{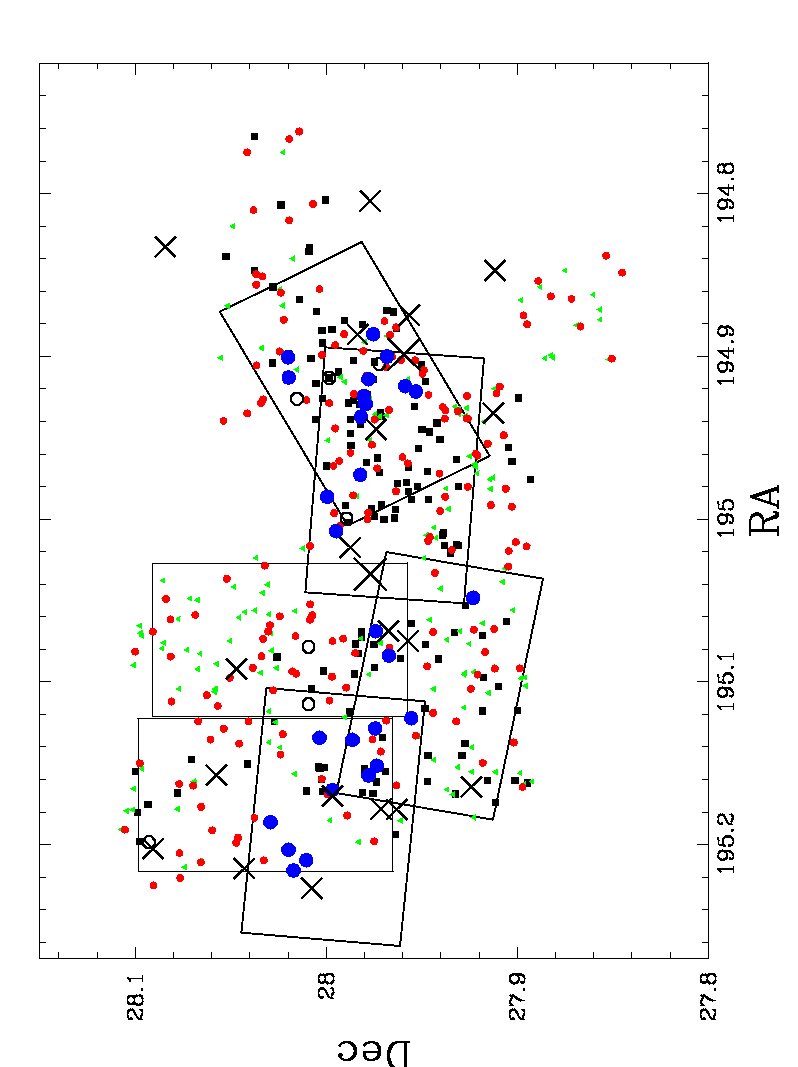}
\caption[]{Maps of confirmed and candidate UCDs (circles and squares,
respectively), cEs (open circles), and giant galaxies (crosses) along with Top: candidate 
globular clusters (points \citep{peng09}) Bottom: confirmed and candidate 
dE (triangles) and dE,N (small circles) Coma members (Trentham et al. in prep).
\label{locall}}
\end{centering}
\end{figure}

The spatial distribution of UCDs are compared to that of the Coma
cluster globular cluster and dE,N populations in Figure \ref{locall}.
Candidate globular clusters \citep{peng09} are mapped as points in the top
plot. There is a clear concentration of globular clusters visible around NGC 4874.  
An overdensity is also apparent surrounding the missing ACS field that would have
included NGC 4889.  
On the eastern side of the cluster, there is a slight overdensity at $\delta = 28.02$ deg near
IC4051 which lies just off our ACS footprint.  Intracluster globular clusters not
associated with individual galaxies are spread throughout the core region.
After smoothing the globular cluster distribution,
\citet{peng09} find an excess which forms a band through the
core between NGC 4874 in the west to IC 4051 and NGC 4908 in the east.  Similarly,
the UCDs confirmed to date exhibit a concentration around NGC 4874 along with
an east-west flattened distribution through the core 
forming a linear structure in the same region as this apparent
band of globular clusters.  Considering candidate UCDs lends further support 
for the presence of a band-like structure running through the core.
However, with the small number statistics for UCDs, it is difficult to tell if this
is a real structure, or if the UCDs are simply following the distribution of more
massive galaxies in the core region.

\begin{figure}[t]
\begin{centering}
\plotone{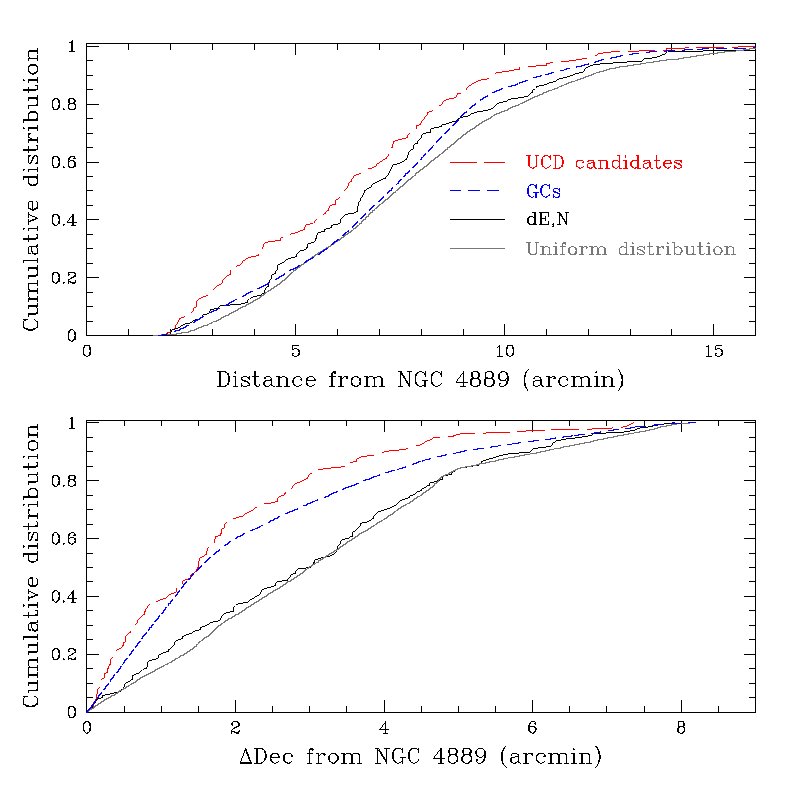}
\caption[]{Cumulative distribution as a function of distance from NGC 4889
out to the extent of the central ACS footprint.  Top: projected
distance from NGC 4889, Bottom: distance in declination only.
The expectation for a uniform distribution (gray solid line) is determined by randomly 
distributing 10,000 points over the observed central ACS fields.
\label{dENdistro}}
\end{centering}
\end{figure}

Meanwhile the dE,Ns exhibit a slight excess near NGC 4874 and NGC 4889, but otherwise 
appear to be spread uniformly
through the ACS footprint, with no indication of either an excess or deficit in the
eastern side of the core where the remainder of the UCDs have been confirmed.
In Figure \ref{dENdistro}, we show the cumulative distribution of confirmed and candidate
UCDs, dE,N, and GCs as a function of projected distance from NGC 4889.  For comparison, we 
take a  sample of 10,000 points distributed randomly over the observed core ACS footprint. 
The UCDs are more centrally concentrated than any other population.  The dE,N are also more 
concentrated than expected for a uniform distribution, while the globular clusters have a 
shallower distribution until $\sim 8$ arcmin in distance from NGC 4889 most likely 
due to the large concentration around NGC 4874.  However, taking the cumulative distributions
as a function of distance in declination only, we find that both GCs and UCDs are much more
concentrated toward the central declination, while dE,N follow the uniform
distribution.  This difference between the populations is greater when considering 
declination only, most likely because of the strong clustering of GCs and UCDs around 
the central giants, and perhaps due to the excess of compact systems noted along a band in 
declination.

In future work, we plan to do more detailed photometric, structural, and distribution
comparisons between Coma UCDs and dE,N galaxies.
For now, we use the luminosity relation found by \citet{cote06} between Virgo dE,N and their 
respective nuclei to compare the magnitudes of UCDs with those expected for the Coma dE,N 
nuclei.  We find that within the Coma footprint, there are only $\sim 20-30$ dE,N within
the magnitude range $15 < F814W < 20$ which could host nuclei with magnitudes in the 
range $21 < F814W < 23.5$.  
We have confirmed 27 UCDs and identified over 100 other candidates.  We 
expect the ratio of UCDs/N ($21 < F814W < 23.5$) to be at least 2-3.  If we assume that all UCDs
are produced from threshing of dE,N, the efficiency of stripping within the Coma core region must be 
very high.  However, given the requirement for highly eccentric orbits, threshing models
do not predict such high efficiency rates \citep{bekki03}.  In this scenario,
one might also expect to find the number of dE,N to be depleted toward the core of the cluster.
Instead, the dE,N are found to have a fairly uniform distribution with perhaps a slight excess
in the central core.  

\begin{figure}[t]
\begin{centering}
\plotone{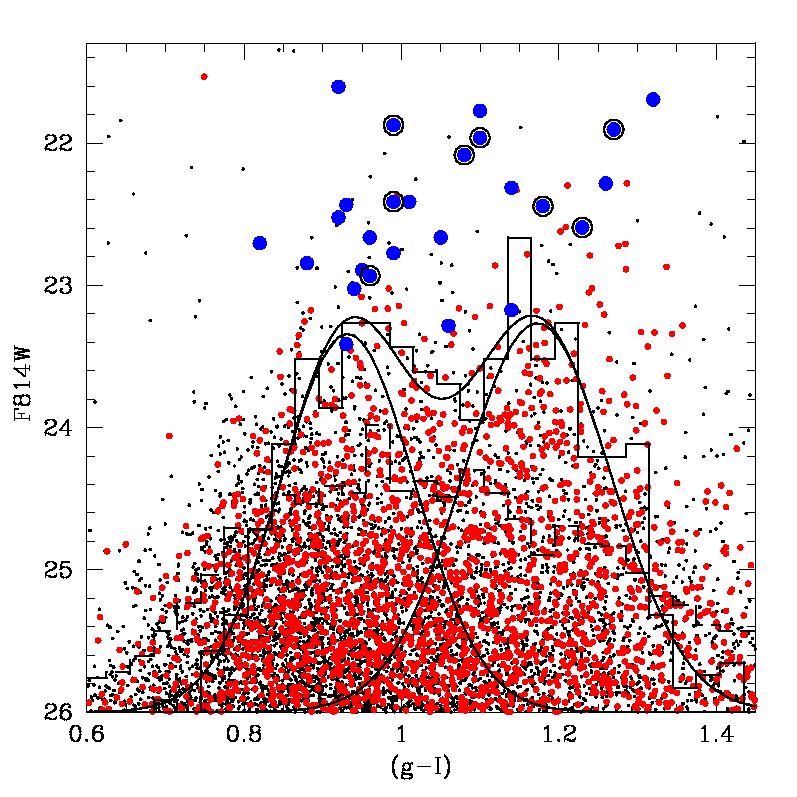}
\caption[]{Color distribution of Coma cluster candidate globular clusters
(points: full ACS survey region, small circles: visit 19 only) and confirmed UCDs (large
circles: all UCDs, encircled: only those from visit 19).
The dashed histogram displays binned $F814W > 24.7$ visit 19 globular cluster
counts, the solid histogram includes only $F814W < 24.7$ visit 19 counts.  We show
the best double Gaussian fit to this set of brighter counts. The
two peaks are at $g - I = 0.93$ and 1.17.
\label{gcol}}
\end{centering}
\end{figure}

The confirmed UCDs display a wide range in color extending to exceptionally red objects.  
One explanation for the extreme red colors is that a dE,N which initially lies along the 
red sequence undergoes threshing and is stripped of at 
least 4 magnitudes of material by the tidal field of the cluster without affecting the color 
of the remnant nucleus. 
We find evidence for an alternative explanation when we compare UCD and globular cluster 
colors.  Globular clusters exhibit a well known bimodal color distribution likely
due to a metallicity bimodality \citep{west}.  This generates a large range in globular 
cluster colors.  In Figure \ref{gcol}, we display the $g - I$ color distribution of the 
confirmed UCDs with that of the Coma cluster globular cluster candidates from \citet{peng09}.
Small circles represent globular clusters from visit 19, the field which includes NGC 4874.  
A histogram of globular clusters fainter than $I > 24.7$ contains only one obvious peak 
around $g-I = 0.95$.
If we look only at visit 19 globular clusters brighter than $I < 24.7$,
the second peak becomes apparent. The histograms in this figure
are scaled to fit within the boundaries of this plot.  The total number
of globular clusters represented by this lower histogram is $\sim500$.
We fit a double Gaussian to the lower histogram and find peaks at
0.93 and 1.17 with $\sigma = 0.09$ for both. For the full sample of GCs,
\citet{peng09} find peaks at $\sim 0.9$ and 1.15.  For inner GCs around NGC 4874
they find a slightly redder blue peak at 0.94.    
The UCDs are found to span the same wide range in color as the GCs.  With
so few confirmed UCDs, it is difficult to determine whether they also 
exhibit a bimodality in color.   To calculate average colors, we assume the
Gaussian distributions for the globular clusters in order to assign a weight to each UCD
for grouping with the red and blue distributions.  We find weighted averages
of blue and red objects of $g - I = 0.95$ and 1.15, very similar to those of
the globular cluster population. Since we are assuming the same distribution, we cannot
argue that they share this bimodality.  However, this does indicate that the UCD distribution
is not inconsistent with that of the globular clusters.  

One secure conclusion that we can draw from this plot is that there is no
discontinuity between the two populations; the UCDs simply extend to brighter
magnitudes while exhibiting the same spread in color as GCs.  In fact, due to the lack
of any discontinuity, the \citet{peng09} globular cluster catalog includes many
of our confirmed and candidate UCDs.  It is worth noting that the
red GC peak is more prominent when considering bright GCs into the range of UCDs
in the field around NGC 4874, where we also find a large fraction of red UCDs.
The possible extension of globular clusters to brighter magnitudes has also been
observed for the cD galaxy NGC 3311 in Hydra where the red  globular cluster sequence 
is seen to extend upward in luminosity into the range of UCDs \citep{wehner08}.

Both NGC 4874 and IC4051 harbor large globular cluster populations, with
S$_N \sim 9.0$ and 12.0, respectively \citep{mfaa01,harris09}.  
We find a large population of UCDs that is almost
certainly bound to NGC 4874 and another group of three UCDs likely to be associated
with IC 4051.  Unfortunately, this latter galaxy lies east of our ACS footprint.
Confirmation of a large UCD population around this non-central, non-cD galaxy would 
lend strong support for a star cluster origin.

We test whether the UCDs can be accommodated by the bright tail of the NGC 4874 globular 
cluster distribution.  Gaussian fits to \citet{peng09} GC candidates
in visit 19 find $I_{peak}=26.2$ with $\sigma=1.2$. However, incompleteness sets
in before the peak.  We therefore take the globular cluster luminosity function (GCLF) 
values found recently for M87 in 
Virgo, $I_{AB,peak}=26.9$ with $\sigma=1.37$ \citep{peng09V,peng09}. This 
produces fewer counts than UCD candidates at the very bright end by $>3\sigma$ 
for $F814W < 22$.  Fainter than $F814W > 22$ ($M_{F14W} > -13$) the UCD 
population is fully consistent with being drawn from a Gaussian distribution of
GCs.  In fact, many of the UCD candidates are included in the GC candidate list.  
We also take the M87 form of the GCLF 
and randomly distribute in magnitude according to this Gaussian distribution
the $\sim 3000$ globular cluster candidates found in the same ACS field as NGC 4874.
From 1000 simulations, we find that the brightest expected globular
cluster is $M_I = -13.0\pm0.4$.  The brightest confirmed
UCD in this field has $M_{F814W} = -13.2$, consistent with the expectations.

\begin{figure}[t]
\begin{centering}
\plotone{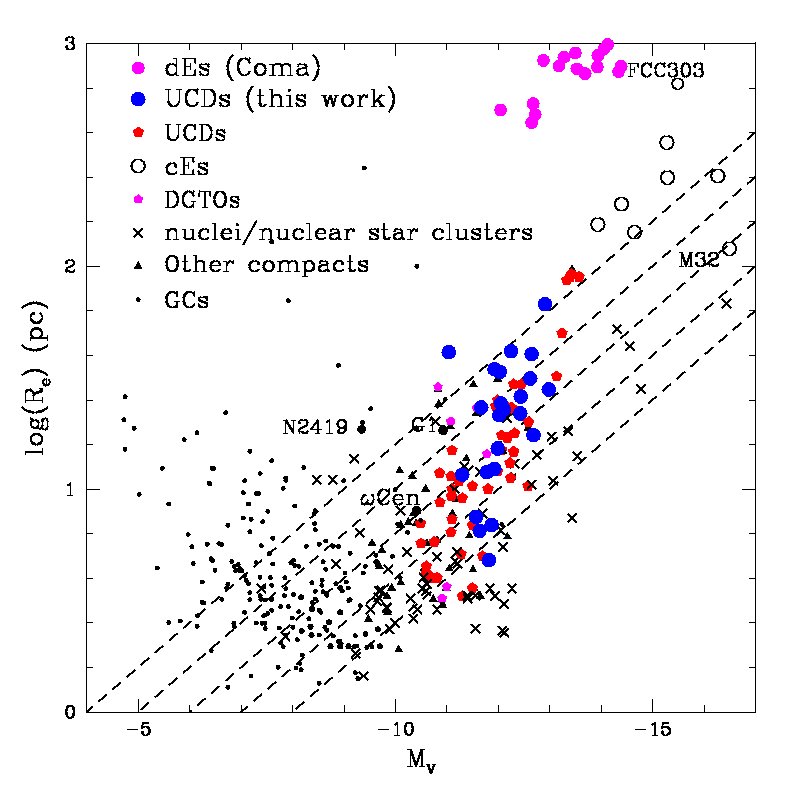}
\caption[]{Magnitude-size relation for compact objects.  Data for UCDs
come from this work and \citet{mieske02,mieske07,mieskeb08,Ev08,hau09}.  Structural
parameters for other object types come from: cEs \citep{compact,Ev08, kent87},
DGTOs \citep{has05}, dE nuclei/nuclear star clusters \citep{geha02,geo09,cote06}, 
other intermediate compact types \citep{mieskeb08,mieske07},
globular clusters \citep{fusi94,grill96,barmby02,mvm05,dacost}, and Coma cluster dEs
\citep{lris,hoyos}.
Dashed lines represent lines of constant surface brightness.  We use the
relation from \citet{peng06} to convert \citet{cote06} $g'$-band data.
\label{sizes}}
\end{centering}
\end{figure}

We compare the sizes of the Coma cluster UCDs with other compact objects in 
Figure \ref{sizes}.  Globular clusters in the range $-10 < M_V < -8$ have 
nearly constant half-light radii $\sim 2.5\pm1.5$ pc independent of luminosity.  
Faintward of this, the observed sizes increase, with effective radii as large 
as 20 pc being found for clusters as faint as $-5$. Brightward of $M_V = -10$,
the few very bright globular clusters start to increase in size with increasing
luminosity.  These objects merge seamlessly into the range of the so-called
UCDs which display a trend of increasing size and decreasing surface brightness 
with increasing luminosity.   The Coma UCDs follow the same luminosity-size relation 
as found for Virgo and Fornax UCDs.  
Compact ellipticals and normal dEs are found toward the bright extension of
this sequence, although cEs would fall off toward higher surface brightnesses while 
dEs have lower surface brightness.
Nuclei of early type galaxies in the Virgo cluster tend to have
smaller sizes at a given magnitude than the UCDs \citep{cote06}. 
This was also noted for nuclear star clusters in low mass dwarf galaxies by 
\citet{geo09}, who suggest that nuclei may expand in size upon being freed from
the strong gravitational potential within a galaxy, e.g. in the case
of galaxy threshing.  

A transition between objects which have sizes independent of magnitude and those which
follow a magnitude - size relation has been previously noted at $M_V \sim -11$.  This 
transition occurs at the same magnitude as a break in the metallicity distribution 
\citep{mieske06b}.  It furthermore corresponds to a mass of 
$\sim 2.5\times 10^{6}$M$_{\odot}$ where 
a break in the $\log \sigma - \log M$ scaling relation has been noted between globular
clusters and larger systems.  Higher mass-to-light ratios, in the range
$6 < M/L_{V} < 9$ have been found for objects above this transition magnitude \citep{has05}.
As these cannot be explained with canonical IMFs and baryonic matter, they suggest
the fundamental difference between globular clusters and UCDs is the presence of
dark matter.  This supports the threshing model in which remnant cores of
threshed dE,N retain some of their dark matter halo.
The nuclei themselves may have formed from the coalescence of globular clusters
through orbital decay within dwarf galaxies \citep{bekki04,lotz}.  This would produce 
complex stellar populations and 
a range of colors/metallicities, along with larger sizes and masses consistent
with what is found for massive globular clusters and perhaps UCDs \citep{geo09,hil06}.   

However, another possibility for the larger sizes and velocity dispersions 
of the UCDs is a difference in physics at the time of formation.
\citet{murray09} shows that star clusters above $M \geq 10^{6} M_{\odot}$ would be optically
thick to IR radiation at the time of formation, and that the balance between radiation
pressure and gravity sets up a size - mass relationship.  If the break in the IMF
is set by the Jeans mass, they argue that optically thick clusters will have top
heavy IMFs.  Top heavy IMFs could be the origin of the high M/L ratios,
since massive stars will age quicker and leave more stellar remnants \citep{dab09}.

Taken together, the properties of the Coma cluster UCDs appear to point toward a star 
cluster origin for a majority of these objects.  The UCDs
exhibit strong spatial and velocity correlations with the major galaxies in the core.  In
particular a large fraction of the UCDs reside within the halos of the massive galaxies, at 
least in the case of the two cD/giant ellipticals in the cluster core. Other UCDs may be 
associated with some of the other giant galaxies in the core region.
The UCDs are furthermore correlated in color and metallicity with the host galaxy.  We
find a large population of red UCDs around NGC 4874 and a bluer population around NGC 4889.
Unfortunately due to the missing ACS field, we do not know whether a radial gradient in 
color/metallicity could be present around NGC 4889.  The range of UCD colors is identical 
to that of globular clusters and there is no evidence for discontinuity between the two 
populations in luminosity.  The spatial distribution shows remarkable similarities to 
globular clusters with a large number of confirmed and candidate UCDs found around the 
same galaxies hosting large globular cluster populations.  Both UCDs and globular clusters 
also show some evidence for a structure of compact objects running 
east-west through the core region.

From the age-metallicity diagram (Figure \ref{metals}), we find that many of the UCDs are not extremely old, 
and thus not primordial in origin.  Red UCDs are more metal rich than blue ones and must
have undergone self-enrichment or been formed from pre-enriched gas.  Red, metal-rich globular
clusters are more often found around massive galaxies \citep{peng06,harris09}.  
The luminosity of globular clusters in a system also correlates with the luminosity of 
the host galaxy, with brighter galaxies hosting populations with brighter
globular clusters \citep{hilker09}.  If UCDs simply extend the globular cluster sequence 
to brighter magnitudes, it then follows that UCDs should be found around the brightest 
and most massive galaxies such as the central Coma cD/giant elliptical galaxies and 
should include a significant number of red objects, as we find.

The commonality in UCD color/metallicity properties with location could be a consequence of 
the formation of UCDs (and GCs) in a small number of discrete star formation events, 
each event characterized by the metallicity of gas available for star formation.  
These star formation events could be induced by 
e.g. cataclysmic wet mergers at early times.  Since each galaxy has a
unique history it follows that the UCD and GC populations should vary from galaxy to galaxy.
The affiliation by color with host suggests
that the UCDs were born in a single event, or only a couple events.  

Some of the blue UCDs are associated with NGC 4889.  Others may follow the globular cluster
intracluster distribution noted by \citet{peng09}.  Some of these blue intracluster globulars
may be metal poor clusters stripped from infalling dwarf galaxies.  In this scenario, it
is possible that the largest
compact stellar systems may be the remnant nuclei of some of these disrupted dwarf galaxies.
However, \citet{peng09} find $\sim 20$\% of intracluster globular clusters are red and
suggest that at least part of this intracluster population may be stripped GCs from the halos
of massive galaxies.  In this scenario, the UCDs could also simply be giant globular 
clusters stripped from these more massive galaxies.  

\begin{figure}[t]
\begin{centering}
\plotone{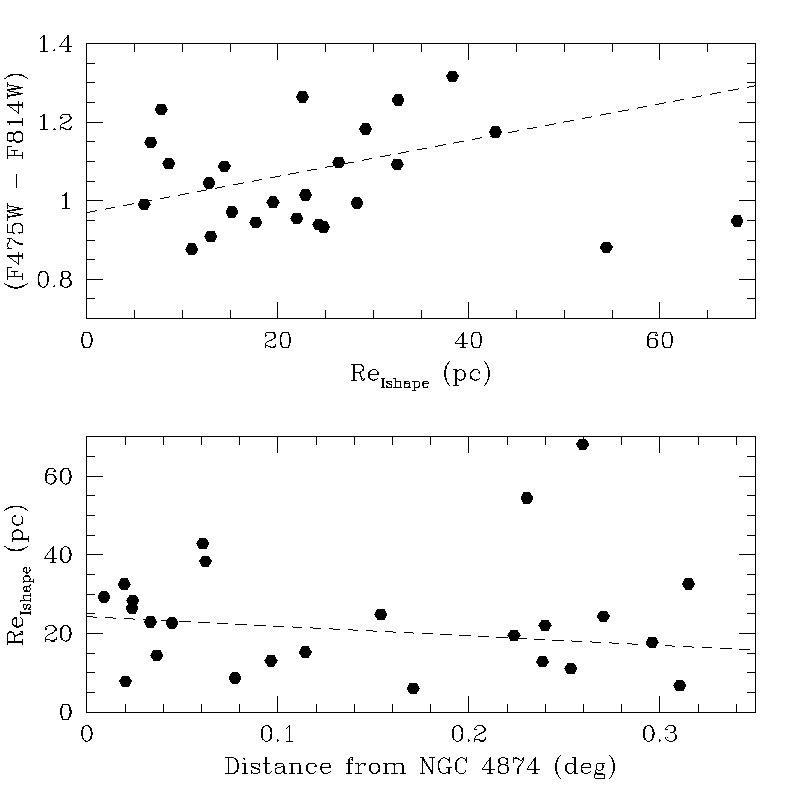}
\caption[]{Size vs color and distance from NGC 4874 vs size for confirmed UCDs.  
Best fit linear relations, excluding two outliers, is shown as the dashed line.
\label{corr}}
\end{centering}
\end{figure}

The findings so far cannot rule out threshing as an origin for UCDs.
The threshing model predicts that UCDs will be found in greater abundance and have
a distribution concentrated toward the center of the cluster potential and around
cluster super giant galaxies \citep{bekki03,bekki07,thomas08}.
This is generally what we find, although detailed modeling would be needed to understand the 
E-W linear structure of compact objects if real.  
The differentiation of color by host is more difficult to explain within the framework of a
threshing scenario.  One possible explanation is if the color variations we find are due to 
radial gradients in metallicity in which more metal rich dE,N are found closer to the 
central giants \citep{mieske06b}, thereby producing an excess of
metal rich UCDs with smaller orbital radii around the central giants.   
Another possibility for explaining the large fraction of red UCDs around NGC 4874 is if more 
massive (and by implication, more metal rich) galaxies require smaller pericenter radii and 
interaction with more massive galaxies for efficient stripping.  Since the sizes of 
dE,N nuclei are known to scale with the brightness of the host dwarf galaxy, we 
would then expect that larger UCDs would be redder and that these would be found 
preferentially close to the central giants.
From Figure \ref{corr} we see only hints of such trends.  After two
outlying UCDs are excluded, we find that larger objects tend to be redder, but with a 
correlation coefficient of only 0.36.  We also find that larger objects tend to be 
closer to NGC 4874, but this trend is also very weak, with a correlation coefficient 
of -0.26.  No correlation is found when comparing UCD size with distance from NGC 4889, 
or with the distance from the closest of either NGC 4889 or NGC 4874, although the 
missing ACS panel centered on NGC 4889 may be the cause of this.

Some similarities between the UCDs and cE galaxies, in particular in the color and 
surface brightness characteristics, are seen.  There may be some correlation in spatial 
distribution as well, although larger numbers of cEs would be needed to establish this.  
Since cE galaxies are expected to form from tidal stripping of more massive galaxies, a 
relation between the two populations would suggest a common threshing origin.  However, we 
have thus far been unable to fill a $\sim 1.5$ mag luminosity gap that currently 
separates these two object types.

We have identified a few confirmed UCDs which may have very faint, low surface 
brightness envelopes.  One of these, object 151072, has obvious surrounding structure, 
the largest measured size of our confirmed UCDs, and a fairly blue color.  Another 
object, 150000, has a hint of a large very low surface brightness envelope and a young derived age.   
Since it is expected that destructive processes such as threshing should play a role in rich 
clusters like Coma it might be expected that at least some fraction of these objects are 
remnants from galactic stripping.  \citet{norris11} make a statistical argument for a bright upper
limit for star cluster formation.  They suggest that compact objects with $M_V > -13$ may be 
a mixture of objects formed through star cluster formation processes and those formed via
a stripping mechanism.  Compact objects brighter than this limit cannot be formed as star
clusters or through the merger of star clusters simply because there are no globular cluster systems 
which are abundant enough to be populated that far into the bright tail of the GCLF.  Object
151072 is our brightest UCD with $M_V = -12.9$, close to this suggested limit.  Two other very bright
UCDs also display hints of surrounding structure.  Object 150000, while not particularly 
bright, does not resemble Coma cluster GCs in stellar population or structural properties. 
Thus, although our findings suggest that the majority of 
Coma cluster UCDs have a star cluster origin, several of the confirmed UCDs which have 
slightly different properties may in fact have 
a galactic origin.  Indeed, a number of other recent studies have also found evidence for
multiple formation channels for UCDs \citep{hil06, darocha10,taylor10,chil10,norris11},
all finding both cases where a star cluster origin is preferred and cases
where stripping of a more massive galaxy is the more likely origin.

The results presented here are based on UCDs located within the core of the Coma cluster.
In future work it will be important to search outer regions as well, to completely
define the spatial and velocity distribution and range of UCD properties.
Future work will discuss the full candidate population and compare in greater detail to 
the properties of Coma cluster dE,N.

\section{Summary and Conclusions}\label{conc}

Using LRIS on Keck, we have spectroscopically confirmed 27 UCDs within the core region of 
the Coma Cluster.
With an initial UCD confirmation success rate greater than 60\% 
for $M_{F814W} < -11.7$, we believe that the rich, dense, evolved environment of the 
Coma Cluster core hosts a large population of UCDs.  However, the high success
rate is also due in part to the fact that UCDs are marginally resolved in ACS imaging.  
We find properties of the UCD population consistent with what has been found for UCDs in
other clusters, having the same range in luminosities, sizes, colors, and metallicities 
and having similar distributions.  

The confirmed Coma cluster UCD population has magnitudes $M_{I} > -13.5$, colors 
$0.8 < (g - I) < 1.3$, sizes primarily in the range $7 - 40$ pc, and 
metallicities $-1.3 \lesssim [Fe/H] \lesssim -0.6$.  They are distributed through the
central core between NGC 4874 in the west and IC 4051 in the east.  We find
strong spatial and velocity correlations with the major cluster galaxies.
A subset of UCDs is almost certainly bound to cD galaxy NGC 4874.  Other 
UCDs may be associated with the other central giant, NGC 4889, and with some of
the other giants in the core region.  Notably, the UCDs also exhibit 
color/metallicity correlations with location in the cluster.  NGC 4874 hosts
a large population of red, metal rich UCDs while NGC 4889 appears to host primarily
blue UCDs, although a radial gradient cannot be ruled out.  There is also a subset 
of blue UCDs which may lie in the intracluster region.   The affiliations with host by color
suggests formation in discrete star formation events with metallicity determined by
that of the gas pool available during such an event, e.g. during the same early time
cataclysmic wet mergers that are believed responsible for producing globular cluster populations.

We suggest that these objects could be related to globular clusters. Not only do they
share similar colors and lie along the extension of the bright tail of the GCLF, 
but they also have a similar distribution in the cluster.   NGC 4874 has a very
high GC specific frequency and hosts a significant UCD population as determined
by the large number of both confirmed and candidate UCDs surrounding 
this galaxy. IC 4051 is another interesting galaxy lying east
of our ACS survey region having an unusually high GC specific frequency and a non-central
location in the cluster. We have identified several UCDs associated with this galaxy and
find a slight excess nearby in our full candidate list.  Confirmation of a large UCD 
population around IC 4051 would provide further evidence for a GC-UCD relationship.
Intracluster GCs have been found distributed throughout the core
region with a possible excess running in an E-W band.  We find a similar structure
with our confirmed UCDs.   Compared to dE,N, possible progenitors in a threshing
scenario, the UCDs are more concentrated
around massive galaxies while dE,N are dispersed through the core. 

Although most UCDs appear to be oversized globular clusters, a few of these objects
exhibit more evidence for a threshing origin.  With hints of diffuse surrounding 
envelopes, bluer colors, younger ages, and/or exceptionally large sizes, these objects may in
fact be remnant nuclei from stripping of dE,N or more massive galaxies.  Since destructive
processes are rampant in dense clusters, it would almost be surprising
if such stripping never occurred.  

A larger sample is required to fully characterize the properties of these objects and probe
the full spatial and velocity distribution of this population.
Establishing the origin of these objects would provide important clues for understanding
galaxy and cluster evolution.
Objects which are remnant nuclei from a previously larger population of dE,N would be 
useful probes for understanding cluster destructive processes, and could help mitigate 
the missing satellite problem.  Objects which prove to be star clusters would be valuable 
probes of galaxy merger and cluster merger and dynamical histories.

\acknowledgments

We thank the anonymous referee for helpful suggestions which have improved this paper.
Based on observations made with the NASA/ESA Hubble Space Telescope, obtained at the Space Telescope
Science Institute, which is operated by the Association of Universities for Research in Astronomy, Inc.,
under NASA contract NAS 5-26555.  These observations are associated with program GO10861.
Support for program GO10861 was provided by NASA through a grant from the Space Telescope Science
Institute, which is operated by the Association of Universities for Research in Astronomy, Inc.,
under NASA contract NAS 5-26555.  Some of the data presented herein were obtained at the W. M.
Keck Observatory, which is operated as a scientific partnership among the California Institute of
Technology, the University of California and the National Aeronautics and Space Administration.
The Observatory was made possible by the generous financial support of the W. M. Keck Foundation.
The authors recognize and acknowledge the very significant cultural role and reverence
that the summit of Mauna Kea has always had within the indigenous Hawaiian community.  We
are most fortunate to have the opportunity to conduct observations from this mountain.

\bibliographystyle{apj}
\bibliography{astr}

\clearpage
\begin{deluxetable}{ccllrrr}
\tabletypesize{\scriptsize}
\tablewidth{0pt}
\tablecaption{Keck/LRIS observations \label{obinfo}}
\tablehead{
\colhead{Mask} &
\colhead{Date} &
\colhead{$\alpha$ (J2000.0)} &
\colhead{$\delta$ (J2000.0)} &
\colhead{PA (deg)} &
\colhead{Seeing (arcsec)} &
\colhead{N$_{obj}$} 
}

\startdata

1 & 2 Apr 2008 & 13 00 40.69 & 28 01 54.86 & 1.5 & 0.7 &  35 \\
2 & 3 Apr 2008 & 13 00 17.82 & 28 01 26.81 & 1.5 & 1.0 &  32 \\
3\tablenotemark{a} & 3 Apr 2008 & 13 00 24.62 & 27 56 26.11 & 80.0 & 1.0 &  38 \\
4 & 2 Apr 2008 & 12 59 40.10 & 27 59 06.28 & 121.0 & 0.8 &  39 \\
5\tablenotemark{b} & 28-29 Mar 2009 & 13 00 43.95  & 27 59 47.08 & 85.0  & 0.8-1.4 & 43  \\
6 & 30 Mar 2009 & 12 59 53.57  & 27 57 50.93  & -94.0  & 1-1.3  & 42  \\

\enddata

\tablenotetext{a}{Exposure times for both red and blue chips were $8\times1500$ sec with the 
exception of Mask 3 for which we obtained only $6\times1500$ sec on the blue side.}
\tablenotetext{b}{Observations were taken in thick and variable cloud cover.}

\end{deluxetable}


\begin{deluxetable}{rrrrrrrrrrrrrrrrrrrr}
\rotate
\tabletypesize{\tiny}
\tablewidth{0pt}
\tablecaption{Targeted UCD candidates\label{ucdtab}}
\tablehead{
\colhead{ID} &
\colhead{$R$} &
\colhead{$F814W$\tablenotemark{a}} &
\colhead{$F814W$} &
\colhead{$B-V$\tablenotemark{b}} &
\colhead{$g-I$\tablenotemark{c}} &
\colhead{$\mu_{\circ,R}$} &
\colhead{$\langle\mu\rangle_{e,F814W}$} &
\colhead{RA} &
\colhead{Dec} &
\colhead{ACS} &
\colhead{x} &
\colhead{y} &
\colhead{cz} &
\colhead{err\tablenotemark{d}} &
\colhead{R$_{fx}$} &
\colhead{SNR\tablenotemark{e}} &
\colhead{cz} &
\colhead{err} \\
\colhead{} &
\colhead{} &
\colhead{corr.} &
\colhead{} &
\colhead{} &
\colhead{} &
\colhead{mag as$^{-2}$} &
\colhead{mag as$^{-2}$} &
\colhead{(J2000.0)} &
\colhead{(J2000.0)} &
\colhead{fld} &
\colhead{} &
\colhead{} &
\colhead{Ab} &
\colhead{km s$^{-1}$} &
\colhead{} &
\colhead{} &
\colhead{Em} &
\colhead{km s$^{-1}$} \\
}

\startdata

\bf{Mem.} &   & & &  & & & & & & & & &  & & & & \\
191006 &   22.88 & 22.48 & 22.60 & 1.03 & 1.28 & 22.58 & 19.08  & 194.8864724 & 27.9753443 & 19 & 2370 &3340 & 7436 & 58 & 9.5 & 11.9 & &  \\
192636 &   21.67 & 22.67 & 22.37 & 0.95 & 1.18 & 22.35 & 19.82  & 194.9001640 & 27.9681118 & 19 & 2699 &2380 & 6929 & 62 &  7.3 & 7.3 &  &    \\
120985 &   22.04 & 22.15 & 22.31 & 0.97 & 1.23 & 22.02 & 19.95  & 194.9007842 & 28.0200830 & 12 & 2965 &3260 & 7578 &59 &  9.8 & 13.2 &  &    \\
121666 &   21.65 & 21.56 & 21.68 & 0.99 & 1.34 & 21.61 & 19.28 & 194.9131763 & 28.0197866 & 12 & 2822 &2484 & 7589 & 56 &  12.5 & 17.8 & &    \\
195526 &   22.12 & 21.82 & 21.95 & 0.84 & 1.10 & 21.72 & 19.50 & 194.9182912 & 27.9585846 & 19 & 3129 &1110 & 6867 & 58 & 10.9 & 14.1 & &    \\
195614 &   22.28 & 22.27 & 22.39 & 0.84 & 1.03 & 22.23 & 19.58 & 194.9245153 & 27.9801757 & 19 & 1526 &1046 & 7366 & 84 &  8.1 & 11.7 & &    \\
196790 &   21.86 & 21.78 & 21.90 & 0.84 & 1.29 & 21.81 & 18.99 & 194.9371432 & 27.9817021 & 19 & 1251 & 284 & 7250 & 53&  16.5 & 14.2 & &    \\
182204 &   23.06 & 23.16 & 23.29 & 0.65 & 1.23 & 23.17 & 20.08 & 194.9727861 & 27.9820669 & 18 & 1615 &2116 & 6470 & 71 & 4.9& 4.7 & &  \\
242857 &   22.38 & 22.30 & 22.43 & 0.65 & 0.95 & 22.19 & 19.57 & 195.0483625 & 27.9229976 & 24 & 1699 & 434 & 6050 & 67 &  6.9 & 13.2  &  &    \\
160141 &   22.91 & 22.65 & 22.78 & 0.63 & 1.04 & 22.51 & 19.31 & 195.0690704 & 27.9741453 & 16 & 2618 &4094 &  6873 & 69 & 5.9 & 5.8 & &    \\
161244 &   22.60 & 22.56 & 22.67 & 0.84 & 1.00 & 22.31 & 19.01 & 195.0841507 & 27.9672299 & 16 & 2906 &3052 &  6290 & 133 &  5.5  & 8.5 &  &    \\
163341 &   22.25 & 22.27 & 22.38 & 0.69 & 1.03 & 22.21 & 19.70 & 195.1224794 & 27.9555062 & 16 & 3224 & 492 &  7017& 86 &  6.9 & 12.7 & &    \\
163400 &   23.42 & 23.27 & 23.40 & 0.68 & 0.91 & 23.25 & 21.26 & 195.1287727 & 27.9744361 & 16 & 1808 & 384 &  6315 & 91 &  3.4 & 5.1 & &    \\
92415  &   22.86 & 22.78 & 22.90 & 0.77 & 0.97 & 22.64 & 20.01 & 195.1343794 & 28.0035249 &  9 & 3620 & 610 &  4681 & 67 &  7.7 & 8.0 & &    \\
163575 &   22.52 & 22.53 & 22.66 & 0.71 & 1.04 & 22.38 & 19.56 & 195.1358599 & 27.9863010 & 16 &  879 & 122 &  6538 & 61 &  8.5 & 9.2 & &    \\
150880 &   22.70 & 22.71 & 22.84 & 0.67& 0.91 & 22.51  & 19.55 & 195.1516839 & 27.9734096 & 15 & 2438 &2988 &  6436 & 77 &  5.1 & 8.0 & &    \\
151072\tablenotemark{f} &  21.48 & 21.47 &  21.60 & 0.71 & 0.96 & 21.65 & 20.14 & 195.1576077 & 27.9779451 & 15 &  2040& 2686&  4906  & 74 &  6.7 & 18.9 & &    \\
 & &  & &  & & &  & & & &  & &  4845  & 59 &  10.0 & 21.3 &  &    \\
150000 &   21.50 &   &    &     0.96 & &   &    & 195.1664874 & 27.9967222 & 15 &  599 &2416 &  6641 & 64 &  9.1 & 11.9 & &  \\
81669 &   23.01 & 22.89 & 23.04 &  & 0.98  & 22.74 & 19.87  & 195.1861561 & 28.0290572 &  8 & 1997 &1815 &  6843& 63 &  7.5 & 7.7 &  &  \\

\bf{Uncert.} &   &  & & & & &  & & & & & &  & & & & \\
190183 &   23.22 & 23.78 & 23.93 & 0.69 & 0.84 & 23.56 & 20.42  & 194.8779144 & 27.9905224 & 19 & 1419 &4102 & 7620 & 152 & 2.1 & 3.3 & &  \\
190875 &   23.38 & 23.53 &  23.69 & 0.75 & 1.05 & 23.62 & 20.63 & 194.8918457 & 28.0019245 & 19 &  432 &3406 & 8977 & 106 & 3.9& 3.3 & &  \\
182306 &   23.46 & 23.37  & 23.49 & 0.67 & 1.07 & 23.31 & 20.08 & 194.9685059 & 27.9567585 & 18 & 3460 &2008 & 6705 & 94 & 3.7 & 4.9 & &  \\
160540 &  22.88 & 23.11 & 23.25 & 0.78 & 0.93 & 22.88 & 19.99 & 195.0771637 & 27.9749966 & 16 & 2444 &3596 &  7014 & 122 &  2.8& 5.0 & &  \\
92668  &  23.52 & 23.34 & 23.48 & 0.90 & 1.07 & 23.31 & 20.02  & 195.1504211 & 28.0412064 & 9  & 759  &184  & 6544 & 93 & 4.3& 1.6 & &  \\
151061 &   23.75 & 23.94 & 24.09 & 0.77 & 0.83 & 23.53 & 20.76 & 195.1549072 & 27.9687271 & 15 & 2732 &2718 & 6819 & 101 & 2.8 & 2.6 & &  \\

\bf{Bckgrd}  & & & &  & &  & & & & & & &  & &  & & & \\
230649 &  22.42 & 22.25 & 22.41 & 0.62 & 1.23 & 22.54 & 21.00 & 195.0646973 & 27.9079971 & 23 & 3408 &3240 &       &    &  &   & 157512 &35  \\
162819 &  23.41 & 23.27 & 23.44 & 0.87 & 0.75 & 23.31 & 21.74 & 195.1097870 & 27.9604759 & 16 & 3035 &1348 & 53558 & 120 & 3.3 & 4.9 & 53837 & 112  \\
221365 &  23.32 & 23.60 & 23.77 & 0.76 & 1.08 & 23.61  & 21.25 & 195.1539307 & 27.9245911 & 22 & 1918 &1984 &        &  &   &      & 161089 & 32  \\
10679  &  23.72 & 23.23 & 23.48 & 0.63 & 0.82 &  23.32 & 22.01  & 195.1783752 & 28.0864925 &  1 & 1998 &3298 &        &  &   &      & 54528 & 34 \\

\bf{Stars}   & & & &  & &  & & & & & & &  & & & &  & \\
180960 &   22.68 & 22.46 & 22.58 & 0.92 & 1.42 & 22.30  & 18.78 & 194.9499969 & 27.9778938 & 18 & 2216 &3474 &  98 & 78 & 4.4 & 6.9 & &  \\
160373 &   21.73 & 21.71 &  21.83 & 0.72 & 1.34 & 21.46  & 17.95 & 195.0763245 & 27.9917908 & 16 & 1272 &3898 &  45 & 56 & 7.7 & 16.4 & &  \\
161617\tablenotemark{f} &   22.80 & 22.68  & 22.81 & 0.75 & 1.05 & 22.49 & 19.05 & 195.0893860 & 27.9666824 & 16 & 2868 &2712 & 149 & 75 & 3.8 & 7.6 & &  \\
 &  &              &      &      &   &        &     &            &   &    &      &     &    69 & 65 &  4.0 & 4.7 & &  \\
221051 &   21.61& 21.69 &   21.80 & 0.59 & 0.57 & 21.25 & 17.98 &  195.1444550 & 27.9183636 & 22 & 2482 &2480 &  -23 & 67 &  4.1 & 14.9 & &  \\
10388  &  21.68 & 21.96 & 22.09 & 1.05 & 2.32 & 21.42 & 19.05 &  195.1722107 & 28.0783367 &  1 & 2654 &3560 &   84 & 53 &  13.0 & 17.0 & &  \\
10355  &   21.56 & 22.01 & 22.14 & 0.84 & 1.94 &  21.39  & 19.24 & 195.1748352 & 28.0928249 &  1 & 1599 &3614 &   58 & 55 &  13.1 & 23.6 & &  \\

\bf{Failed}   & &  & & & & & & &  & & & &  & & & &  & \\
196829& 23.51& 23.81 & 23.93 & 0.57& 0.84 & 23.71 & 20.33 &  194.9307098 & 27.9536591 & 19 & 3317&  266 & \\
123628& 23.99& 23.97 & 24.14& 1.07& 1.38 & 23.92 & 21.18 &  194.9386597 & 28.0055618 & 12 & 3476&  684 &  \\
181622 &  23.93 & 23.88 & 24.09 & 0.66 & 0.83 &23.75  & 21.41 & 194.9625397 & 27.9730034 & 18 & 2395 &2622 &  & &  & & & & \\
181520& 23.49& 23.43 & 23.58& 0.80& 1.17 & 23.23 & 19.99 & 194.9673767 & 27.9999008 & 18 &  437& 2724 & \\
182263& 23.58 & 23.92 & 24.07& 0.56 & 0.73 & 23.77& 20.96 & 194.9713745 & 27.9721527 & 18 & 2338& 2060 &  \\
230603& 23.72 &23.72 & 23.87& 0.52& 0.94 & 23.46 & 21.04 &  195.0628357 & 27.9055920 &23  &3602 &3320  & \\
160317& 23.61 & 23.96 & 24.11& 0.80& 0.85 & 23.88 & 20.42 &  195.0726318 & 27.9816227 & 16 & 2037& 3976 & \\
161324 & 23.60 & 23.62 & 23.75 & 1.00 & 1.09 & 23.40  & 20.30 & 195.0933685 & 28.0015354 & 16 &  361 &2986 &  & & & &  &  &\\
232327& 23.31 & 23.35 & 23.50& 1.02& 1.40 & 22.99& 20.11 &  195.1030426 & 27.9097061 &23  &2780 & 878  & \\
221570& 23.70 & 23.80 & 23.99& 0.52& 1.23 & 23.60 & 22.09 &  195.1613312 & 27.9467297 &22  & 261 &1854  & \\
151353& 23.71 &24.13 & 24.27& 0.80& 0.91 & 23.90 & 21.08 &  195.1680908 & 27.9807396 &15  &1712 &2078  &  \\
81133 &  & 23.83  & 23.99&     & 2.54  &   & 21.34 &  195.1739960 & 28.0168438 & 8  & 3023& 2396 &  \\

\enddata

\tablenotetext{a}{Kron magnitudes corrected for light loss due
to the finite size of the Kron aperture as compared to the ACS PSF \citep{hammer}.}
\tablenotetext{b}{Measured within a 3.0 arcsec aperture.}
\tablenotetext{c}{Measured within a 2.25 arcsec aperture.}
\tablenotetext{d}{Includes measurement uncertainty and uncertainty in wavelength calibration
shifts based on sky lines.}
\tablenotetext{e}{Signal-to-noise ratio per Angstrom around 5000\AA.}
\tablenotetext{f}{Object was observed in two masks.}
                                                                                                                                           
\end{deluxetable}

\begin{deluxetable}{rrrrrrrrrrrrrrrrrrrr}
\rotate
\tabletypesize{\tiny}
\tablewidth{0pt}
\tablecaption{UCD Sample 2\label{ucdtab2}}
\tablehead{
\colhead{ID} &
\colhead{$R$} &
\colhead{$F814W$\tablenotemark{a}} &
\colhead{$F814W$} &
\colhead{$B-V$\tablenotemark{b}} &
\colhead{$g-I$\tablenotemark{c}} &
\colhead{$\mu_{\circ,R}$} &
\colhead{$\langle\mu\rangle_{e,F814W}$} &
\colhead{RA} &
\colhead{Dec} &
\colhead{ACS} &
\colhead{x} &
\colhead{y} &
\colhead{cz} &
\colhead{err} &
\colhead{R$_{fx}$} &
\colhead{SNR\tablenotemark{d}} &
\colhead{cz} &
\colhead{err} \\
\colhead{} &
\colhead{} &
\colhead{corr.} &
\colhead{} &
\colhead{} &
\colhead{} &
\colhead{mag as$^{-2}$} &
\colhead{mag as$^{-2}$} &
\colhead{(J2000.0)} &
\colhead{(J2000.0)} &
\colhead{fld} &
\colhead{} &
\colhead{} &
\colhead{Ab} &
\colhead{km s$^{-1}$} &
\colhead{} &
\colhead{} &
\colhead{Em} &
\colhead{km s$^{-1}$} \\
}

\startdata

\bf{Mem.} &   & & &  & & & & & & & & &  & & & & \\
1041346 & 21.65 & 21.77 & 21.89 & 0.87 & 0.89 & 21.71 & 19.34 & 194.9141693 & 27.9777908 & 19 & 1837 & 1658 & 8816 & 40 & 6.41 & 9.1 & &  \\
1039188 & 21.97 & 21.97 & 22.10 & 1.47  & 1.00 & 22.00 & 19.33  & 194.9216614 & 27.9530296 & 19 & 3481 & 819 & 6955 & 40  & 8.40 & 6.3 & & \\
1041508 & 22.62 & 22.82 & 22.95 & 0.71 & 0.83 & 22.62 & 19.89  & 194.9293823 & 27.9793549 & 19 & 1525  & 735 & 7346 & 66 & 3.63 & 3.9 & & \\
1043225 &  & 22.61 & 22.72 &  & 0.65 &  & 20.07  & 194.9863739 & 27.9994297 & 18 & 218 & 1536 & 6950 & 49 & 3.68 & 3.9 & & \\
1042830 & 21.83 & 21.67 & 21.79 & 0.80 & 1.02  & 21.59 & 18.79  & 195.0075073 & 27.9948120 & 18 & 264 & 152 & 9438 & 39 & 6.97 & 9.2  & & \\
2000005 & 23.19 & 23.05 & 23.19 & 0.50 & 1.04 & 23.19 & 19.67 & 195.2032471 & 28.0197811 & 8 & 2430 & 623 & 7339 & 40 & 7.92 & 3.0 & & \\
1044251 & 22.31 & 22.15 & 22.30 & 0.77 & 1.19 & 22.24 & 19.73 & 195.2095490 & 28.0104160 & 8 & 3005 & 89 & 4563 & 35 & 8.61 & 6.3  & & \\
1044847 & 22.59 &  &  & 0.88  &  & 22.49  &   & 195.2160492 & 28.0170689 &  &  & & 8489 & 50 & 5.54 & 4.2 & & \\

\bf{Uncert.} &   &  & & & & &  & & & & & &  & & & & \\
1039715 & 22.65 & 22.61 & 22.69 & 0.79 & 1.10 & 22.75 & 20.49 & 194.9122467 & 27.9591923 & 19 & 3170 & 1498 & 6558 & 70 & 3.10 & 4.7 & & \\
1041400 & 22.78 & 22.59 & 22.73 & 0.92 & 1.19 & 22.61 & 19.79 & 194.9237061 & 27.9782658 & 19 & 1677 & 1072 & 6982  & 54 & 5.19 & 2.7 & & \\
1038942 & 22.94 & 22.96 & 23.11 & 0.64 & 0.79 & 23.11 & 20.73 & 194.9502506 & 27.9502506 & 25 & 181 & 2693 &6538  & 154 & 2.56 & 3.6 & & \\
1039079 & 22.70 & 22.74 & 22.88 & 0.74 & 0.99 & 22.69 & 19.53 & 194.9569092 & 27.9517746 & 18 & 3964 & 2655 & 9638  & 68 & 5.40 & 3.4 & & \\
1037297 & 22.89 & 22.90 & 23.04 & 0.64 & 0.92 & 22.79 & 19.88 & 194.9636688 & 27.9317131 & 25 & 1355 & 1797 & 7153 & 139 & 3.14 & 2.8 & & \\
1040263 & 22.91 & 22.70 & 22.81 & 0.83 & 0.84 & 22.81 & 21.02 & 194.9937897 & 27.9655056 & 18 & 2509 & 565 & 6821 & 124 & 3.49 & 3.4 & & \\
1038505 & 23.10 & 22.97 & 23.12 & 0.43 & 0.72 & 22.69 & 20.45 & 195.0224152 & 27.9456463 & 24 & 456 & 2388 &5777  & 72 & 3.83 & 3.3 & & \\
1045724 & 22.32 & 22.39 & 22.51 & 0.97 & 1.27 & 22.18 & 18.89 & 195.1241608 & 28.0267639 & 9 & 2123 & 1600 &6896 & 54 & 4.76 & 4.8 & & \\

\bf{Bckgrd}  & & & &  & &  & & & & & & &  & &  & & & \\
1037801 & 21.17 & 21.16 & 21.27 & 0.45 & 0.43 & 21.36 & 20.07 & 194.9686737 & 27.9382534 & 25 & 828 & 1585 &  44368 & 129 & 2.40 & 17.5 & 44454 &6  \\
1042772 & 23.22 & 23.23 & 23.37 & 0.65 & 0.98 & 23.41 & 21.52 & 194.9895020 & 27.9941273 & 18 & 551 & 1262 &    &   &    &    & 122313 & 87  \\
1042434 & 23.49 &  &  & 0.43 &  & 23.27 &  & 195.0279236 & 27.9900494 &  &  & &    &   &    &    & 110080  & 76  \\
1041404 & 21.43 & 21.36 & 21.49 & 0.70 & 0.44 & 21.56 & 20.45 & 195.1437988 & 27.9785004 & 15 & 2191 & 3556 &    &   &    &    & 53247  & 36  \\
1040963 & 23.25 & 23.35 & 23.51 & 0.82 & 0.98 & 23.26 & 21.36 & 195.1675415 & 27.9730701 & 15 & 2260 &  1999 &   &   &    &   & 106399  & 63  \\
1042863 &  & 22.61 & 22.77 &  & 0.47 &  & 21.24 & 195.1869965 & 27.9949245 & 15 & 463 & 1115 &       &    & &   & 53043  &50   \\
4000025 &  & 24.62 &  &  &  &  &  & 195.1993713 & 27.9944324 & 15 & 334 & 339 &  81397 & 90 & 2.72 & 2.5  & 81556 & 60  \\

\bf{Stars}   & & & &  & &  & & & & & & &  & & & &  & \\
1039346 & 22.21 & 22.14 & 22.25 & 0.76 & 0.90 & 21.93 & 18.45 & 194.9500885 & 27.9550133 & 18 & 3827 & 3128 & 2  & 93 & 3.42 & 6.5 & & \\
1039823 & 21.06 & 21.32 & 21.44 & 0.64 & 0.29 & 20.82 & 18.14 & 194.9639130 & 27.9605064 & 18 & 3257 & 2350 &    -43 & 25 & 11.4 & 16.4 & &\\
1044234 & 21.82 & 21.79 & 21.91 & 0.82 & 1.08 & 21.60 & 18.19 & 195.1475372 & 28.0103035 & 8 & 3833 & 3943 &   -114 &32 & 9.00 & 10.2 & &\\
1043666 & 22.04 & 22.07 & 22.19 & 0.67 & 0.64 & 21.68 & 18.51 & 195.1602478 & 28.0043449 & 8 & 4085 & 3064 &    -200 &28 &9.38 & 11.8 & &\\
1045115 & 21.62 &  &  & 0.54  &  & 21.36  &  & 195.2292023 & 28.0200977 &  &  & &    -139 & 43 & 6.41 & 15.7 & &\\
8039100 & 22.52 &  &  & 0.80  &  & 22.22  &  & 195.2524567 & 27.9879341 &  &  & &    185 & 44 & 5.53& 5.8 & & \\


\enddata

\tablenotetext{a}{Kron magnitudes corrected for light loss due
to the finite size of the Kron aperture as compared to the ACS PSF \citep{hammer}.}
\tablenotetext{b}{Measured within a 3.0 arcsec aperture.}
\tablenotetext{c}{Measured within a 2.25 arcsec aperture.}
\tablenotetext{d}{Signal-to-noise ratio per Angstrom around 5000\AA}
                                                                                                                                           
\end{deluxetable}

\begin{deluxetable}{rrrrrrrrrr}
\tabletypesize{\scriptsize}
\tablewidth{0pt}
\tablecaption{Effective Radii\label{ucdsize}}
\tablehead{
\colhead{}  &
\colhead{GALFIT}  &
\colhead{(Sersic)}  &
\colhead{}  &
\colhead{}  &
\colhead{}  &
\colhead{ISHAPE}  &
\colhead{}  &
\colhead{}  &
\colhead{} \\
\colhead{ID} &
\colhead{R$_{e}$ (pc)} &
\colhead{n} &
\colhead{$F814W$} &
\colhead{b/a\tablenotemark{b}} &
\colhead{}  &
\colhead{profile} &
\colhead{index} &
\colhead{R$_{e}$ (pc)} &
\colhead{b/a\tablenotemark{b}} \\
}
                                                                                  
\startdata

191006 &  $7.1 \pm 2.3$  &  5.2  & 22.51 & 0.82 &  & King & 30 & $7.8\pm2.6$ & 0.68 \\
192636 & $33.3 \pm 3.8$  &  3.1  & 22.28 & 0.86 & &  King & 30 & $29.2\pm6.5$ & 0.93 \\
120985 & $40.5 \pm 3.2$  &  1.7  & 22.17 & 0.90 & &  King & 30 & $42.8\pm12.2$ & 0.88 \\
121666 & $42.0 \pm 13.3$  &  3.9  & 21.54 & 0.99 & &  King & 100 & $38.3\pm3.6$ & 0.88 \\
195526 & $37.6 \pm 14.9$ & 5.7   & 21.66 & 0.91 &  & King & 100 & $32.5\pm5.3$ & 0.99 \\
195614 & $22.0 \pm 2.5$  &  6.9  & 22.23 & 0.84 &  & King & 100 & $22.9\pm5.2$ & 0.89 \\
196790 & $20.9 \pm 2.4$  &  6.6  & 21.74 & 0.86 &  & King & 100 & $22.6\pm6.1$ & 0.87 \\
182204 & $14.7 \pm 4.0$  &  5.4  & 23.17 & 0.65 &  & King & 100 & $8.6\pm4.6$ & 0.91 \\
242857 & $23.7 \pm 3.0$  &  4.7  & 22.30 & 0.79 &  & King & 30 & $24.8\pm5.2$ & 0.79 \\
160141 &  $7.0 \pm 3.1$  &  5.1  & 22.66 & 0.76 &  & King & 100 & $6.0\pm2.9$ & 0.89 \\
161244 &  $4.8 \pm 2.9$  &  3.7  & 22.60 & 1.00 & &  & &  &  \\
163341 & $23.3 \pm 2.7$  &  7.6  & 22.24 & 0.98 &  & King & 100 & $19.5\pm6.7$ & 0.87 \\
163400 & $66.9 \pm 16.4$  &  6.3  & 23.11 & 0.74 &  & King & 100 & $54.4\pm7.1$ & 0.80 \\
 92415 & $24.3 \pm 3.8$  &  7.6  & 22.67 & 0.91 &  & King & 100 & $22.0\pm6.4$ & 0.87 \\
163575 & $17.1 \pm 4.1$  &  5.3  & 22.53 & 0.74 &  & King & 100 & $12.8\pm6.6$ & 0.90 \\
150880 & $12.8 \pm 5.9$  &  4.4  & 22.74 & 0.95 &  & King & 100 & $11.0\pm2.4$ & 0.77 \\
151072 & $125.5 \pm 18.8$ &  6.6  & 21.29 & 0.98 &  & King & 100 & $68.1\pm6.4$ & 0.89  \\
150000 & $36.0 \pm 8.3$  &  7.9  & 22.38 & 0.92 &  & King & 100 & $24.3\pm2.8$ & 0.72 \\
 81669 & $18.2 \pm 6.0$  &  2.6  & 22.95 & 0.95 &  & King & 30 & $17.7\pm5.4$ & 0.97 \\
1041346 & $27.7 \pm 7.5$ &  4.4  & 21.89 & 0.94 &  & King & 100 & $28.3\pm4.5$ & 0.85 \\
1039188 & $25.6 \pm 4.2$ &  5.2  & 21.97 & 0.89 &  & King & 100 & $26.4\pm6.0$ & 0.88 \\
1041508 & $10.2 \pm 3.9$ &  6.1 &  22.91 & 0.74 &  & King & 100 & $14.4\pm4.6$ & 0.86 \\
1043225 & $20.9 \pm 7.0$ &  4.2 &  22.92 & 0.64 &  & King & 100 & $13.0\pm4.2$ & 0.74 \\
1042830 & $19.7 \pm 4.8$ &  4.1 &  21.87 & 0.97 &  & King & 100 & $15.2\pm6.8$ & 0.94 \\
2000005 & $7.1 \pm 3.3$ &  4.1 &  23.07 & 0.85 &  & King & 100 & $6.7\pm2.5$ & 0.98 \\
1044251 & $34.7 \pm 8.4$ & 4.5 & 22.15 & 0.84  &   & King & 30 & $32.6\pm6.7$ & 0.85 \\
1044847\tablenotemark{a} &                &     &        &      &   &      &     &              &      \\

\enddata

\tablenotetext{a}{This object is outside of our ACS coverage.  Although this object has been observed in
archival WFPC2 images, the larger pixel scale makes size measurements for such
small objects impossible.}
\tablenotetext{b}{Typical uncertainties in b/a were 0.07 and 0.05 for Galfit and Ishape, respectively. However, for
such small objects, we don't consider the axial ratio measurements to be very reliable.}

\end{deluxetable}

\begin{deluxetable}{rrrrrrrrrr}
\tabletypesize{\scriptsize}
\tablewidth{0pt}
\tablecaption{Age and Metallicity\label{Tmetals}}
\tablehead{
\colhead{Type}  &
\colhead{N}  &
\colhead{H$\beta$}  &
\colhead{Fe 5270}  &
\colhead{Mgb}  &
\colhead{Age}  &
\colhead{[Fe/H]}  &
\colhead{[Mg/Fe]}  &
\colhead{$V - I$\tablenotemark{a}} &
\colhead{[Fe/H]$_{(V-I)}$\tablenotemark{b}} \\
\colhead{} &
\colhead{} &
\colhead{\AA} &
\colhead{\AA} &
\colhead{\AA} &
\colhead{Gyr} &
\colhead{dex} &
\colhead{dex} &
\colhead{mag} &
\colhead{dex} \\
}
                                                                                  
\startdata

UCDs- red & 6 &  $1.53\pm0.23$ & $2.40\pm0.32$ & $2.42\pm0.29$ & $14.5^{max}_{5.4}$ & $-0.62^{max}_{0.22}$ & $-0.10^{0.11}_{0.12}$ &  1.18 & $-0.41$ \\
UCDs- blue & 9 & $2.41\pm0.31$ & $0.99\pm0.47 $ & $1.33\pm0.36$ &  & &  & 0.95 & $-1.40$  \\
UCDs- bright &5 &  $2.01\pm0.34$ & $1.84\pm0.45$& $2.73\pm0.40$ & $8.7^{max}_{4.6}$  & $-0.80^{0.42}_{0.41}$ & $0.44^{0.51}_{0.51}$  \\
UCDs- red,N4874& 5  & $1.73\pm0.25$  & $2.22\pm0.33$ & $2.60\pm0.31$ & $12.6^{max}_{4.6}$  & $-0.64^{0.26}_{0.26}$ & $0.14^{0.23}_{0.23}$ & 1.18 & $-0.40$   \\
UCDs- strong H$\beta$ & 8 & $2.99\pm0.28$ & $1.09\pm0.41$ & $1.98\pm0.32$ & $4.0^{1.0}_{0.7}$  & $-1.24^{max}_{0.57}$  & $0.76^{0.30}_{0.27}$  & 0.97 & $-1.31$ \\
UCDs- weak H$\beta$ & 8 & $1.20\pm0.28$ & $1.92\pm0.39$& $2.02\pm0.39$ & &  & & 1.08 & $-0.84$ \\
UCDs- metal rich & 2 & $1.63\pm0.43$ & $3.08\pm0.54$ & $2.52\pm0.52$ & $13.2^{max}_{7.7}$ & $-0.13^{0.31}_{0.26}$ & $-0.32^{0.21}_{0.20}$   & 1.17 & $-0.47$ \\
UCD/dE,N & 2 & $2.48\pm0.22$  & $0.76\pm0.31$ & $1.03\pm0.27$ & $5.2^{1.5}_{1.1}$  & $-1.02^{0.24}_{0.12}$ & $-0.07^{0.18}_{0.18}$ & 1.07 & $-0.87$ \\
dE,N(brt) & 4 & $2.26\pm0.19$  & $1.85\pm0.27$ & $1.28\pm0.24$  & $6.0^{1.9}_{1.6}$  & $-0.73^{0.23}_{0.23}$ & $-0.13^{0.15}_{0.16}$ & 1.08 & $-0.82$ \\
dE,N(fnt) & 24 & $2.42\pm0.14$  & $1.84\pm0.20$ & $1.04\pm0.17$ & $4.6^{1.2}_{1.0}$ & $-0.66^{0.17}_{0.16}$ & $-0.24^{0.10}_{0.10}$  & 1.06 & $-0.91$   \\
dEs & 10 & $2.19\pm0.29$  & $1.29\pm0.39$ & $0.82\pm0.33$ & $7.4^{3.6}_{2.0}$ & $-1.16^{max}_{0.29}$ &$-0.18^{0.21}_{0.19}$   \\
 &  &  &   &  & &  \\
Individual: &  &  &   &  & &  \\
150000 &  & $3.01\pm0.62$  & $1.57\pm0.85$ & $1.82\pm0.75$ & $2.1^{2.2}_{0.7}$ & $-0.40^{0.60}_{max}$ & $0.04^{0.30}_{0.20}$ &   &  \\
196790 &  & $1.95\pm0.66$  & $2.71\pm0.91$ & $4.18\pm0.83$ & $7.3^{max}_{4.5}$ & $-0.08^{0.34}_{0.25}$ & $0.24^{0.27}_{0.27}$  & 1.25 & $-0.12$  \\
121666 &  & $1.84\pm0.44$  & $3.48\pm0.57$ & $4.27\pm0.51$ & $7.6^{max}_{max}$ & $0.17^{0.16}_{max}$ & $0.02^{0.12}_{0.12}$ & 1.17 & $-0.47$  \\
cE (91543) &  & $2.08\pm0.12$  & $3.04\pm0.15$ & $4.22\pm0.13$ & $4.8^{1.4}_{1.3}$ & $0.08^{0.07}_{0.09}$ & $0.20^{0.09}_{0.08}$ & 1.33 & 0.22  \\

\enddata

\tablenotetext{a}{Average values are used for combined spectra.}
\tablenotetext{b}{Estimated from the relation between ($V-I$) color and [Fe/H] provided
by \citet{barmby00}}
\tablecomments{Where age and metallicity values are absent, this is due to measured line indices falling 
outside of model ranges.}

\end{deluxetable}

\begin{deluxetable}{lrrrrr}
\tabletypesize{\scriptsize}
\tablewidth{0pt}
\tablecaption{Velocity Distribution\label{ucdvels}}
\tablehead{
\colhead{Sample}  &
\colhead{Number}  &
\colhead{$\langle v_r \rangle$}  &
\colhead{err}  &
\colhead{$\sigma_v$}  &
\colhead{err} \\
\colhead{} &
\colhead{} &
\colhead{km s$^{-1}$} &
\colhead{} &
\colhead{km s$^{-1}$} &
\colhead{} \\
}
                                                                                  
\startdata

Full LSB  & 51 &  6970  & 178 & 1269 & 126  \\
Full UCD  & 27  &  6887  & 207 & 1072 & 146  \\
$(V-I) > 1.05$ & 11 & 6992  & 379 & 1260  & 269 \\
$(V-I) > 1.10$ & 8 & 6907  & 325 & 922  & 231 \\
$(V-I) < 1.05$ & 14 & 6707 & 231 & 861  & 163 \\
$\alpha < 195$ & 12 & 7296 & 160 & 558 & 114 \\
$\alpha > 195$ & 15 & 6559 & 325 & 1257 & 230 \\
N4874 proximity &  9 & 7257 & 87 & 254 & 60 \\
N4889 proximity & 5 & 6526 & 148 & 332 & 105 \\

\enddata

\end{deluxetable}


\clearpage
                                                                                  
\clearpage

\clearpage

\clearpage

\clearpage

\clearpage

\clearpage

\clearpage
                                                                                      
\clearpage

\clearpage

\clearpage

\end{document}